\documentclass[a4paper,10pt]{article}
\usepackage{amsmath,amsfonts}
\usepackage{graphicx}

\font\msytw=msbm9 scaled\magstep1

\let\a=\alpha \let\b=\beta \let\g=\gamma \let\d=\delta
\let\e=\varepsilon \let\z=\zeta  
\let\l=\lambda \let\m=\mu \let\n=\nu  
\let\s=\sigma \let\t=\tau \let\f=\varphi 
   
\let\D=\Delta \let\L=\Lambda  
\let\Si=\Sigma   
\let\ee=\epsilon \let\r=\rho \let\th=\theta \let\io=\infty

\def\ie{{i.e. }}\def\eg{{e.g. }}

\def\EE{{\cal E}}

  \def\OO{{\cal O}}

 \def\xx{{\bf x}}

\def\to{\rightarrow} \def\la{\left\langle} \def\ra{\right\rangle}

\def\RRR{\hbox{\msytw R}}

\newcommand{\beq}{\begin{equation}} \newcommand{\eeq}{\end{equation}}
 \newcommand{\wt}{\widetilde}

\begin{document}

\title
{Is it possible to experimentally verify the fluctuation relation? \\
A review of theoretical motivations and numerical evidence
}

\author{Francesco Zamponi \\
{\it Service de Physique Th\'eorique, 
Orme des Merisiers, 
CEA Saclay,} \\ {\it 91191 Gif sur Yvette Cedex, France}
}

\maketitle

\tableofcontents

\newpage

\section*{Foreword}
\addcontentsline{toc}{section}{Foreword}

The statistical mechanics of nonequilibrium stationary states of dissipative systems
and, in particular,
the large deviations of some specific observables attracted a lot of interest in
the past decade. The literature on this problem is enourmous and it is impossible
to give here a comprehensive list of references.

Part of this interest concentrated on the {\it fluctuation relation}, which is a
simple symmetry property of the large deviations of the entropy production rate
(related to the dissipated power). The fluctuation relation is a parameterless
relation and is conjectured to hold in some generality.
The discovery of this relation \cite{ECM93} 
motivated many studies: and experiments specifically designed to its test have been
reported on turbulent
hydrodynamic flows \cite{CL98,CGHLPR04,BCG06} and Rayleigh-B\'enard convection \cite{STX05},
on liquid crystals \cite{GGK01}, on a resistor \cite{GC05},  
on granular gases \cite{FM04}.

Unfortunately, despite this great experimental effort, the situation is very confused,
more than ten years after \cite{ECM93}.
After many debates,
numerical simulations established that the relation holds very generally for
{\it reversible} chaotic dissipative systems: while experiments gave promising results but revealed
also some difficulties in the interpretation of the data that generated many controversies.

Indeed, the apparent generality of the fluctuation relation led to
the idea that it could
be tested ``blindly'' just by measuring the fluctuations of the injected power
in some dissipative system. On the contrary, experience revealed that to test this relation one has
to face many difficulties, and that each experiment has to be interpreted by considering
its own specificities. 

Moreover, when applying the theory to a real physical system, one should obviously keep
in mind that the real system does {\it not} coincide exactly with the mathematical model
we use to describe it. So that, even if we expect that the theory will work for the mathematical
model, its application to the description of the experiment might require further 
efforts.

The aim of this paper is to review some of these difficulties, discovered in more than one
decade of trials, in the hope that new experiments will be performed allowing to clarify
many aspects that are still poorly understood.

However, before turning to this discussion, it is important to review the theoretical developments
that generated the current interest in the fluctuation relation. Indeed, this relation is the most
accessible prediction
of a much more deep theory, and discussing this theory in some detail will be very useful for the 
interpretation of the experiments.

I will focus on the stationary state (or Gallavotti-Cohen) fluctuation relation. There are
other fluctuation relations, such as the (Evans-Searles) transient fluctuation relation,
the Jarzynski equality, and the Crooks fluctuation relation, which hold in much more generality.
These relations have found many interesting applications in different fields, for instance
biophysics of large molecules. However they are not directly related to the ideas
discussed here. A review of these relations is beyond the scope of this paper.

Everywhere in the paper I will only discuss the main ideas, giving references to the original
literature for more detailed discussions.

\section{Why should we test the fluctuation relation?}
\label{sec:introduction}

\subsection{Definitions}

To begin, we will fix some notations. We will consider a dynamical system
described by a set of state variables $x = (x_1,\cdots,x_N)$ and 
evolving according to the equation of motion $\dot x = F(x)$. A trajectory
of the system generated by an initial datum $x_0$ will be indicated by
$x(t)$, while segments of trajectory of duration $\t$ will be indicated
by $\xx(t)$, \ie $\xx(t) = \{x(t), t \in [t_0,t_0+\t]$. Given an 
observable which is a function of the state $x$, $O(x)$, we will indicate
the average over a segment of trajectory by
\beq
O[\xx(t)] = \frac1\t \int_{t_0}^{t_0+\t} dt \, O(x(t)) \ ,
\eeq
and sometimes we will use the shorthand notations $O(t) \equiv O(x(t))$, and
$O_\t \equiv O[\xx(t)]$, so that
\beq
O_\t = \frac1\t \int_{t_0}^{t_0+\t} dt \, O(t) \ .
\eeq

\subsection{The fluctuation relation}

The {\it fluctuation relation} was first introduced in \cite{ECM93}. There, a
 system of Lennard-Jones--like particles subject to a dissipative force 
(inducing {\it shear flow}) and to a thermostatting mechanism was
 investigated. The aim was to test the conjecture that the nonequilibrium 
stationary state reached by the system after a transient could be described by
 a probability distribution over the segments of trajectory $\xx(t)$ 
having the following form\footnote{The expression (\ref{SRBrozza})
has to be intended in the following sense: if one is able to generate $M$ segments
$\xx_i(t), i=1,\cdots,M$ 
of trajectory of duration $\t$ {\it with uniform probability}, then the average
of an observable $O[\xx(t)]$ which is a functional of the trajectory can be computed
as $\sum_{i=1}^M O[\xx_i(t)] \mu\{ \xx_i(t)\} / \sum_{i=1}^M \mu\{ \xx_i(t)\}$. 
However this is not the case in an
experiment: in this case the segments will be generated already according to the weight
(\ref{SRBrozza}) and the average of an observable must be computed, as usual, 
as $M^{-1} \sum_{i=1} O[\xx_i(t)]$. This distinction was not clear in \cite{ECM93}
and this generated some confusion in the subsequent literature.}:
\beq\label{SRBrozza}
\mu\{ \xx(t) \} \sim \frac{\L_u^{-1}[\xx(t)]}{Z} 
\sim \frac{e^{- \t \sum_{i,+} \l_i[\xx(t)]} }{Z } \ .
\eeq
In the latter expression, $\L_u[\xx(t)]$ is the expansion factor over the trajectory,
 related to the sum of all
 {\it positive} Lyapunov exponents computed on the segment $\xx(t)$ (see
 Appendix \ref{app:defi} for a precise definition of all these quantities). The
 normalization factor is $Z = \sum_{\xx(t)} \L_u^{-1}[\xx(t)]$.
 The proposal (\ref{SRBrozza}) originated from earlier
 studies of chaos and turbulence~\cite{ER85} and from periodic orbit 
expansions~\cite{CvBook}

Clearly a discretization in the ``space of trajectories'' $\xx(t)$ is needed to give
a precise mathematical sense to the expression (\ref{SRBrozza}). This problem
is highly non trivial:
a theory leading to an invariant measure of the form (\ref{SRBrozza}), based
on {\it Markov partitions} to discretize phase space, was developed
in \cite{Si68a,Si68b,BR75,Ru76,Si77,Ru78,GC95a,GC95b,Ga95b}, see also \cite{Ru95,Ru99,Ga00,Ga02,GBG04} 
and in particular \cite{Ru04,Ga95} for less technical discussions
and \cite{Ga06} for a nice recent review of this problem. In this context
the measure (\ref{SRBrozza}), seen as a measure on phase space\footnote{This is done
by relating trajectories to initial data generating them, after a Markov partition
of phase space has been constructed. In particular, $x = \xx(0)$ is the point
in the middle of the segment $\xx(t) \, , \, t \in [-\t/2,\t/2]$ 
and the limit $\t \to \io$ has to be properly taken.} $\m(x)$,
is called Sinai-Ruelle-Bowen (SRB) measure.

The system studied in \cite{ECM93} was described by {\it reversible} equations
of motion. Then, it is straightforward to show \cite{ECM93,GC95a,GC95b,Ga95b} 
that, if $I\xx(t)$ is the time-reversed
of $\xx(t)$,
\beq\label{probratio}
\frac{\mu\{ \xx(t) \} }{\mu\{ I\xx(t) \}} = \frac{e^{- \t \sum_{i,+} \l_i[\xx(t)]} }
{e^{- \t \sum_{i,+} \l_i[I \xx(t)]}} = \frac{e^{- \t \sum_{i,+} \l_i[\xx(t)]} }
{e^{ \t \sum_{i,-} \l_i[ \xx(t)]}} =
e^{-\t \sum_i \l_i[\xx(t)] } \equiv e^{\t
    \s[\xx(t)]} \ ,
\eeq
using the symmetry properties of the Lyapunov exponents under time-reversal,
\ie $\sum_{i,+} \l_i[I\xx(t)] = -\sum_{i,-} \l_i[\xx(t)]$, see Appendix~\ref{app:defi}.

In (\ref{probratio})
the {\it phase space contraction rate} $\s[\xx(t)] = -\sum_{i} \l_i[ \xx(t)]$, equal
to minus the sum of all Lyapunov exponents, appeared. This quantity can
be easily measured in a numerical simulation, as it is possible to show 
(Appendix~\ref{app:defi}) that
\beq
\s[\xx(t)] = \frac1\t \int_0^\t dt \, \s(x(t)) \ , \hskip30pt 
\s(x) = -\nabla \cdot F(x) \ ,
\eeq
\ie $\s(x)$ is minus the divergence of the right hand side of the equation of motion.
Its average is $\s_+ \equiv \la \s(x)\ra = \la
\s[\xx(t)]\ra$. An infinitesimal volume $dx$ will evolve according to
$\frac{d}{dt}dx = -\s(x) dx$.

It is possible to compute the
probability of observing a value $\s[\xx(t)] = \s$, $P(\s) = P\{ \s[\xx(t)]=\s \}$. 
Using (\ref{probratio}) and $\s[I\xx(t)]=-\s[\xx(t)]$, we get
\beq\label{FRrozza}
P(\s) = \sum_{\xx(t) | \s[\xx(t)]=\s} \m\{\xx(t)\} =
\sum_{\xx(t) | \s[\xx(t)]=\s} e^{\t \s} \m\{I\xx(t)\} =
e^{\t \s} \sum_{\xx(t) | \s[\xx(t)]=-\s} \m\{\xx(t)\} = e^{\t \s} P(-\s) \ .
\eeq
The time-reversibility of the dynamics and the assumption
(\ref{SRBrozza}) imply that $P(\s)$ should verify the symmetry relation 
(\ref{FRrozza}), which is a first example of a fluctuation
  relation and was observed to be true in the numerical simulation 
of~\cite{ECM93}.

Some remarks are in order at this point. \\
{\it 1.} Eq.~(\ref{SRBrozza}), and consequently (\ref{FRrozza}), are the leading contributions 
to the probability for $\t
\to \io$, but for finite $\t$ corrections are present. \\
{\it 2.} As it will be
discussed in detail in the following, if $\s[\xx(t)]$ is unbounded a more
complicated analysis is required; thus it is convenient for the moment to restrict the
attention to systems such that $\s[\xx(t)]$ is bounded. \\
{\it 3.} It can be proven that $\s_+ \geq 0$ under very general
hypotheses \cite{Ru96}. Moreover, if $\s_+=0$ the SRB measure (\ref{SRBrozza}) reduces
to the volume measure\footnote{Or eventually to a measure which is dense with respect
to the volume, $\m(dx) = \r(x)dx$, \eg the Gibbs distribution.} 
 and the system is at equilibrium. The relation (\ref{FRrozza})
reduces to $P(\s)=P(-\s)$. Thus in the following we will assume that $\s_+ > 0$ and in
this case the system will be called {\it dissipative} \cite{Ru04,Ga06}. In this case volumes
will contract on average and the SRB measure will be concentrated on a set of zero volume in
phase space.

Under these hypotheses, defining the
normalized variable $p = \s[\xx(t)]/\s_+$, we expect that for large $\t$ the
probability of $p$ will be described by a {\it large deviation function}, \ie
that\footnote{By $o(\t)$ we mean a quantity $Q_\t$ such that $Q_\t/\t \to 0$ for $\t\to\io$.}
\beq\label{zdef}
P(p \s_+) \sim e^{\t \z_\io(p) + o(\t)} \ .
\eeq
As $\s$ is bounded, $p$ is also bounded and for $\t \to \io$ we have
$|p|<p^*$; the value of $p^*$ can be easily identified as we expect that
$\z_\io(p) = -\io$ for $|p|> p^*$, \ie deviations bigger than $p^*$ have zero
probability\footnote{Note that the equality 
$p^* = \max{|\s(x)|}/\s_+$ is {\it not} true in general; obviously
$p^* \leq  \max{|\s(x)|}/\s_+$ is verified, but in general the value of $p^*$
is smaller because the observation of $p = \max{|\s(x)|}/\s_+$ over a very long
trajectory requires that $\s(x(t)) \equiv \max{|\s(x)|}$ for all $t$ along the trajectory,
which is clearly very unlikely and might have zero probability for large $\t$.} 
in the limit $\t \to \io$.

Using (\ref{FRrozza}) and (\ref{zdef}) we get the following relation \cite{GC95a,GC95b,Ga95b}:

\vskip5pt
\noindent
{\it Defining the large deviation function $\z_\io(p) = \lim_{\t\to\io} \frac1\t
  \log[P(\s[\xx(t)]=p \s_+)]$, and $p^*$ as the maximal value of $|p|$ for which
  $\z_\io(p) > -\io$, we have}
\beq\label{FR}
\z_\io(p) - \z_\io(-p) = p \s_+ \ , \hskip20pt |p| < p^* \ .
\eeq

We will refer to the proposition above as the {\it fluctuation relation}.
The two conditions that $\t$ is large and that $p$ is bounded are often
misconsidered but are very important and should not be forgotten \cite{Ga95b}
if one wants to avoid erroneous interpretations of (\ref{FR}).

By looking at the sketchy derivation in (\ref{FRrozza}), one realizes that the
fluctuation relation can be extended in the following way. Consider an observable
$O[\xx(t)]$ that is even or odd under time reversal, \ie $O[\xx(t)]=\pm O[I\xx(t)]$, and
the normalized variable $q = O[\xx(t)]/O_+$, where $O_+=\langle O[\xx(t)] \rangle$ is
the average of $O$ in the stationary state.
Consider the joint probability $P(p,q) \sim \exp \t \z_\io(p,q)$. One can show
\cite{Ga96a,Ga96b} that 
\beq\label{FRestesa}
\z_\io(p,q) - \z_\io(-p,\pm q) = p \s_+ \ , \hskip20pt |p| < p^* \ , \ |q|<q^* \ ,
\eeq
and the sign in (\ref{FRestesa}) must be chosen according to the sign of $O$ under
time reversal. Eq.~(\ref{FRestesa}) can be extended to any given number of observables
having definite (eventually different) parity under time reversal.

The fluctuation relation attracted a lot of interest after \cite{ECM93}
because it is a parameter-free relation that might hold in some generality in
nonequilibrium systems. For this reason in the last decade it has been
numerically tested
on a lot of different models, often with positive result, and sometimes with
negative or confusing results. In experiments the situation is complicated by
the presence of many noise sources and by the difficulty to find a good
modellization of the system under investigation.

One should keep in mind that, at least in this context
\footnote{As discussed in the foreword, the study
of different fluctuation relations 
is beyond the scope of this paper.},
the main theoretical motivation to study the fluctuation relation, 
that was already at the basis of the work \cite{ECM93}, 
is to obtain some insight on the measure
describing stationary states of nonequilibrium system. 
This is clearly a very large class and it is possible to find nonequilibrium
systems displaying any kind of strange behavior.
We wish to identify a class of nonequilibrium systems such that
the fluctuation relation holds in their stationary states.

\subsection{The ``transient fluctuation relation''}

Subsequently after \cite{ECM93}, it was noted that an apparently similar fluctuation
relation holds in great generality. Namely, we can consider, instead of
segments of trajectory drawn from the stationary state distribution,
segments originating from initial data extracted from an equilibrium
distribution (\eg the microcanonical one). In other words, we extract initial
data according to an equilibrium distribution and then evolve them using the
dissipative equation of motion. If the equations of motion are reversible, it
is possible to show that, if $\xx(t) \, , t\in [0,\t]$ is the trajectory
starting from $x_0=x(0)$ extracted with the equilibrium distribution one 
has~\cite{ES94}
\beq\label{TFT}
\frac{P_{eq}\{\s[\xx(t)]=\s\}}{P_{eq}\{\s[\xx(t)]=-\s\}} = e^{\t \s} \ .
\eeq
The latter relation has been called {\it transient fluctuation relation} and
is very easy to prove: it follows from the definition of $\s$, see Appendix~\ref{app:TFT}.
It holds in great generality, not only for the microcanonical ensemble but for
many other equilibrium ensembles, see \cite{ES02} for a review\footnote{To avoid confusion
note that the point of view on this subject
expressed in \cite{ES02} and in other papers by Evans and coworkers
is very different from the one expressed here.}. The fundamental difference
between (\ref{TFT}) and (\ref{FR}) is that in the former trajectories are
sampled according to the equilibrium distribution of their initial data, while
in the latter they are sampled according to the nonequilibrium stationary state 
distribution\footnote{It is interesting to remark that the phase space contraction
rate is defined with respect to a given measure on phase space, see Appendix \ref{app:defi}. 
The fluctuation relation
(\ref{FR}), as an asymptotic statement for large $\t$, holds independently of the chosen measure 
(at least for smooth systems, see 
section \ref{sec:singularities}), while the transient fluctuation relation (\ref{TFT})
holds for finite $\t$ only if the phase space contraction rate is defined with respect to the
equilibrium measure from which one extracts initial data.}.

Then one can ask whether the fluctuation relation (\ref{FR}) can be derived
starting from (\ref{TFT}). Naively one could take
the limit $\t \to \io$ of (\ref{TFT}), claim that the initial transient is negligible
and assume that (\ref{TFT}) holds also for the stationary distribution.
However, this is not at all trivial. Depending on the properties of the
system, the probability $P_{eq}$ might not converge to a well defined
probability distribution in the limit $\t\to\io$, or the convergence time might
be very large (practically infinite), and so on. 

Even if $\t^{-1} \log P_{eq}(p\s_+)$ converges fast
  enough to a smooth limiting function $\wt\z_\io(p)$, which will verify the
  fluctuation relation by (\ref{TFT}), the latter may be different from the true function
$\z_\io(p)$ describing the stationary state~\cite{GC99}: 
indeed the measure (\ref{SRBrozza}) is concentrated on a set of zero volume in phase space
if $\s_+ >0$, and making statements on a set of zero measure starting from a set of initial 
data extracted according to the volume measure might be very difficult.
We will give examples in section~\ref{sec:time_scales}.

  Understanding in full generality
  what are the conditions that allow to extract the fluctuation relation (\ref{FR})
from the transient relation (\ref{TFT}) seems to be a difficult task and is for
the moment an unsolved problem. 
To simplify the problem and try to understand its main
  features, we can, following \cite{GC95a,GC95b}, restrict our attention to a
  class of simple dissipative systems for which the fluctuation relation can be 
rigorously proved whithout making reference to (\ref{TFT}), but directly from
the invariant measure (\ref{SRBrozza}).

\begin{figure}[t]
\centering
\includegraphics[width=.7\textwidth]{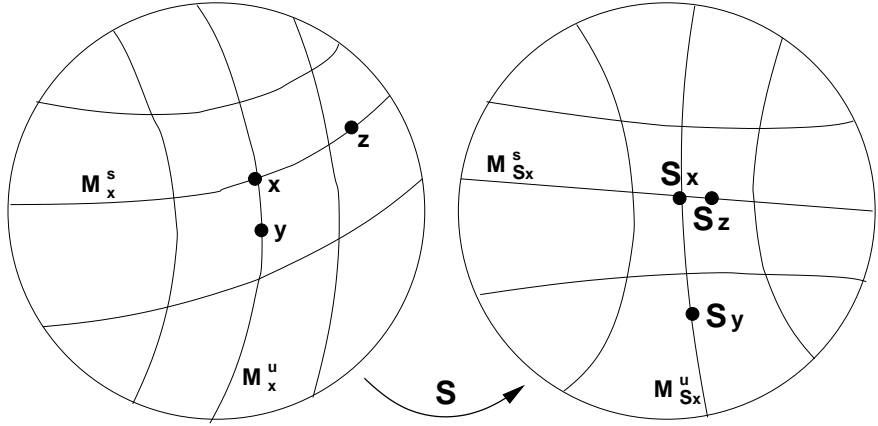}
\caption[Pictorial representation of an Anosov system]{
A pictorial representation of an Anosov system. In the vicinity of a point $x \in M$
it is possible to draw two families of manifolds $M^{s,u}$. The manifolds passing
through $x$ are the stable and unstable manifolds of $x$, $M^{s,u}_x$. 
Under the action of $S$, the manifolds $M^{s,u}_x$ are mapped into the manifolds
$M^{s,u}_{Sx}$ passing through $Sx$. A point $y \in M^u_x$ is mapped into a point
$Sy$ whose distance from $Sx$ is larger by a factor $\sim \L$, while a point 
$z \in M^s_x$
is mapped into a point $Sz$ which is closer to $Sx$ by the same factor $\L$.
}
\label{fig:anosov}
\end{figure}

\subsection{Anosov systems and the fluctuation theorem: a proof of the
  fluctuation relation}

In \cite{GC95a,GC95b} a proof of the fluctuation relation was given
for Anosov systems. The rigorous mathematical proof for Anosov maps
(discrete time) is in \cite{Ga95b}, while the proof for Anosov flows
(continuous time) has been given in \cite{Ge98}.

Anosov systems are paradigms of chaotic systems. A precise mathematical definition
is in Appendix~\ref{app:anosov}. Roughly speaking, they are defined by
a compact manifold $M$ (phase space) and a smooth (\ie at least twice
differentiable) map $S(x)$ acting on $x \in M$ (the dynamics), having the
following properties, see Fig.~\ref{fig:anosov} for an illustration:

\vskip5pt
\noindent
{\it (1)} Around each $x \in M$ it is possible to draw two smooth surfaces
$M^u_x$ and $M^s_x$ such that points $y \in M^s_x$ approach $x$
exponentially while points $y \in M^u_x$ diverge
exponentially from $x$ under the action of $S$ ({\it existence of the stable and 
unstable manifolds});

\vskip5pt
\noindent
{\it (2)} the rate of convergence (divergence) on the stable (unstable)
manifold is bounded uniformly on $M$ ({\it uniform hyperbolicity});

\vskip5pt
\noindent
{\it (3)} the surfaces $M^{s,u}_x$ vary continuously w.r.t. $x$, and
the angle between them is not vanishing; under the action of $S$ the surfaces
$M^{s,u}_x$ are mapped into the corresponding $M^{s,u}_{Sx}$
({\it smoothness of the stable and unstable manifolds});

\vskip5pt
\noindent
{\it (4)} there is a point $x$ which has a dense orbit in $M$ ({\it the
  attractor is dense}). In this case the system is called {\it transitive}.

\vskip5pt
In particular, {\it 1)} ensures that the system is chaotic;
{\it 2), 3)} ensure that the system is smooth enough so that the
measure (\ref{SRBrozza}) is well defined; and {\it 4)} ensures that the
attractor is dense on $M$ so that there are no regions of finite volume that
do not contain points visited by the system in stationary state.
Note that as $M$ is compact and $S(x)$ is smooth, it follows that $\s[\xx(t)]$
is smooth and bounded, as we assumed above.

For Anosov systems it can be shown, by explicitly constructing a Markov partition
\cite{Si68a,Si68b,Si77,GBG04}, that, if the initial
data are extracted from the uniform distribution over $M$ (\ie the
microcanonical distribution), the system reaches exponentially fast a
stationary state described by the measure (\ref{SRBrozza}), which is called
Sinai-Ruelle-Bowen (SRB) measure. The function $\z_\io(p)$ exists and it is analytic
and convex for $p_1 < p <p_2$~\cite{Si68a,Si68b,Si77}. If the system is transitive, dissipative
($\s_+>0$) and
reversible, $-p_1=p_2=p^*$ and $\z_\io(p)$ verifies the fluctuation
relation~\cite{GC95a,GC95b} in the form (\ref{FR}). This is the 
{\it Gallavotti-Cohen fluctuation theorem}.

Moreover, in this case the function $\t^{-1} \log P_{eq}(p \s_+)$ converges
to the limiting large deviation function $\z_\io(p)$ describing the stationary state
and satisfying the fluctuation theorem.

\subsection{Stochastic systems (in brief)}

Alternatively, one can consider a stochastic model for a dissipative
system. This was first done by Kurchan \cite{Ku98}, who considered a Langevin system
in presence of a dissipative force. He showed that Eq.~(\ref{TFT}) holds also
in this case for any finite time $\t$, and studied the conditions under which
it converges to a well defined limiting distribution. It turns out again that
the main requirements are some smoothness properties (see also \cite{PRV06}
for a discussion) and the existence of a gap in the spectrum of the Fokker-Planck
operator. The latter requirement implies that the probability distribution
describing the system converges exponentially fast to the equilibrium
distribution and that correlation functions decay exponentially, and can be
considered as the counterpart of chaoticity in Langevin systems.
In the context of stochastic systems, the fluctuation relation was also
derived by Lebowitz and Spohn for discrete Markov processes~\cite{LS99}.
Extensions and different perspectives 
were given by Maes~\cite{Ma99}, Crooks~\cite{Cr99}, Bertini et al.~\cite{BDGJL01,BDGJL05},
Derrida et al.~\cite{DLS02}, Depken~\cite{De02}.
The discussion of the stochastic case would require much more space but is beyond the scope
of this paper.

\subsection{The chaotic hypothesis: a class of nonequilibrium stationary states verifying the fluctuation relation}
\label{sec:ingredients}

The general features that are behind all the existing derivations of the
fluctuation relation are the following:

\vskip5pt
\noindent
{\it 1)} {\sc Reversibility}: the equations of motion should be
reversible. This is the symmetry that is at the basis of the fluctuation
relation, as it is clear from the derivation in (\ref{probratio}), (\ref{FRrozza});

\vskip5pt
\noindent
{\it 2)} {\sc Smoothness}: the system must be smooth enough, and in particular
the phase space contraction rate is assumed to be smooth and bounded; this
ensures that $p$ is well defined\footnote{Note that an obvious requirement
is that $\s_+ > 0$, \ie that the system is {\it dissipative}; otherwise $p$
is not defined and the fluctuation relation reduces to the trivial identity
$P(\s)=P(-\s)$. The limit $\s_+ \to 0$ is non trivial as we will see in
the following.} and bounded and that it exists a value $p^*$
such that $\z_\io(p)=-\io$ for $|p|>p^*$;

\vskip5pt
\noindent
{\it 3)} {\sc Chaoticity}: the system must be chaotic in the sense of having
at least one positive Lyapunov exponent (for deterministic systems) or of
having a gap in the spectrum of the Fokker-Planck operator (for Langevin
systems); chaoticity implies that the stationary state is
reached exponentially fast and that the function $\t^{-1} \log P_{eq}(p \s_+)$
converges to a limiting distribution $\z_\io(p)$;

\vskip5pt
\noindent
{\it 4)} ``{\sc Ergodicity}'' (or transitivity): by this we mean that the attractor should be
dense in phase space, \ie that the system must be able to visit any finite
volume of its phase space in stationary state. The reason why this property is
important will be discussed below. This property can be checked, for instance, by
looking at the Lyapunov spectrum, see \cite{BG97,Ga99} and references therein.

\vskip5pt

All these ingredients seem to be important for the fluctuation relation to
hold. It is worth to remark that properties {\it 1)-4)} cannot, obviously, be {\it directly}
tested in an experiment. As often in physics, we are
here making hypotheses on the mathematical properties of a {\it model} that we assume 
is able to describe the system under investigation.
From these hypotheses, we then derive some consequences, in the form of relations between observables 
(\eg the fluctuation relation), and only the latter are accessible to the experiment.

As far as I know, for all investigated models satisfying {\it 1)-4)} the
fluctuation relation has been succesfully verified. On the other hand,
examples are known of systems violating at least one of the conditions above
and for which the fluctuation relation does not hold. If one of the
requirements above is violated a case-to-case
analysis is needed.

However, it is likely that experimental systems are described by models 
that violate some of the requirements above, in particular the smoothness and 
ergodicity requirements. It is then
interesting to investigate in more detail what happens in these cases to see
if, under less restrictive hypotheses, we can still draw some general
conclusions on how the fluctuation relation will be modified.

In this context, a {\it chaotic hypothesis} has been proposed by Gallavotti
and Cohen \cite{GC95a,GC95b}: it states that, even if hypothesis {\it 1)-4)} cannot be proven to
be satisfied in a strict mathematical sense,
{\it for the purpose of computing the averages of some particular observables 
of physical interest, still
the system can be thought as an Anosov system and its invariant measure can be assumed to be
given by Eq.(\ref{SRBrozza})}.

It is worth to stress that, even if we accept the chaotic hypothesis, we
should take care in drawing consequences from it. The violation of one of the
hypotheses above will be observed if one chooses a suitable observable, \eg by
probing motion in extreme regimes (for example, looking at {\it very} large
deviations of $\s$, in a sense to be made precise below). Thus, in applying
the chaotic hypothesis to a given system, one must take into account its
peculiarities to avoid contradictions, in the same way one uses the ergodic
hypothesis for equilibrium systems: see \cite{Ga06} for a review.

In the following, we begin by reviewing the evidence in favor of the validity
of the fluctuation relation for system that do not violate
hypothesis {\it 1)-4)} in a substantial way; at the same time we will discuss
some difficulties that are encountered when trying to experimentally verify
the fluctuation relation.
Then we will discuss some classes of systems in which one among {\it 1)-4)} is
not true and discuss what might happen in these cases.

\section{Verification of the fluctuation relation in reversible, smooth and chaotic systems}
\label{sec:GK}

First of all, we wish to discuss the difficulties that are {\it intrinsically} present
if one wishes to test the fluctuation relation. At the end of this section,
we will review the numerical simulations that attempted to verify the fluctuation
relation in systems that are not proven to satisfy requirements {\it 1)-4)} of section
\ref{sec:ingredients}, but at least do not seem to violate
them in a substantial way.

So let us assume for the moment that the system under investigation verifies the hypotheses of
the fluctuation
theorem (for instance, it is a reversible Anosov system), but we want to test the
fluctuation relation numerically (to debug our program).

The main problem is that to test Eq.~(\ref{FR}) we need to observe
(many) negative fluctuations of $\s_\t$ for large $\t$. To be precise, one should
construct the function $\z_\t(p) = \frac1\t \log P(p\s_+)$ and check that {\it it is
independent of $\t$ in a given interval of $p$}. If this interval contains $p=0$, one 
can test the fluctuation
relation in this interval. This is very important to guarantee that preasymptotic
effects can be neglected.

In general, $\s_\t$ will be proportional to the number
of degrees of freedom (it is extensive).
The system has to be {\it dissipative}, \ie the average $\s_+$ must be positive,
otherwise the fluctuation relation is trivial. This means that the maximum of $P(\s_\t)$
(or equivalently of the function $\z_\t(p)$ defined above)
will be assumed close to $\s_\t = \s_+$ (or $p=\s_\t/\s_+=1$). The fact that $\s_\t$ is extensive 
implies that in general $\z_\io(p) \propto N$, where $N$ is the number of degrees of 
freedom in the system.

The function $\z_\io(p)$ being convex, 
the probability to observe a negative value of $p$ will be smaller than the probabity
of $p=0$. Thus we can estimate this probability as
\beq
P[\text{negative fluctuation}] \sim e^{\t \z_\io(0)} \ .
\eeq
The value of $\z_\io(0) < 0$, thus this probability will be very small as long as $\t$
is large. Moreover, $\z_\io(0) \propto N$ so the probability will scale as $\exp(-N\t)$.
In general, we have\footnote{See the discussion after Eq.~(\ref{zsmallE}) and Appendix A in 
\cite{ZBCK05} for an explicit computation in a very simple case.}, for small $\s_+$,
$\z_\io(0) \sim \s_+ = \t_0^{-1} N \s_0$, where $\t_0$ is the microscopic characteristic time
of the system and $\s_0$ is the (adimensional) phase space contraction per degree of freedom
over a time $\t_0$. Finally we have
\beq\label{pnegstima}
P[\text{negative fluctuation}] \sim e^{-(\t/\t_0) N \s_0} \ .
\eeq

\subsection{Entropy production rate}

In many models of dissipative systems it turns out,
by explicit computation, that the phase space contraction rate is given by the power
dissipated into the system divided by its temperature, \ie it can be identified with
an {\it entropy production rate} \cite{Ga06b}. This is crucial to
define an experimentally observable counterpart of the phase space contraction
rate, if one wants to test the fluctuation relation.

The theoretical discussion of this identification is beyond the scope of this paper.
So we will take a more practical point of view. Assume, as in section \ref{sec:ingredients},
that we are able to build a model that we believe is able to describe the experimental
system we wish to investigate. We can then compute the phase space contraction rate for
this particular model. If this quantity turns out to be measurable in the 
experiment, we can use it to perform a test of the fluctuation relation.
Otherwise, if the phase space contraction rate does not correspond to an observable quantity,
we cannot use this system to test the fluctuation relation. However, this seems not to be
the case at least for the models that have been considered in the literature: it always turned
out that the phase space contraction rate could be identified with a measurable entropy
production rate \cite{Ga06b}.

Accepting this identification, we can now give a quantitative estimate of the probability
in (\ref{pnegstima}) in a physical example.
Consider an experiment done 
on a resistor; we apply a field $E$ and measure the current $J$ flowing trough the resistor.
The dissipated power is $E J$ and the entropy production is $E J /k_B T$, where $T$ is the
temperature of the resistor. We average this quantity over a time $\t$ to obtain
\beq\label{entropyav}
\s_\t = \frac{1}{\t k_B T} \int_0^\t dt E J(t) \ .
\eeq
It is clear that the probability of observing a spontaneous reversal of the current
must be {\it very} small. Let us estimate it using equation (\ref{pnegstima}). 
We consider a resistor with $R=1 \ k\Omega$, a current $I = 1 \ mA$, and an observation time
$\t = 1 \ \m s$, at temperature $T = 300 K$ and with a microscopic time (\ie the time
between two collisions of electrons in the resistor) $\t_0 \sim 10^{-13} s$. 
We get $\t/\t_0 = 10^7$ and using $k_B T \sim 10^{-21} \ J$ we get
$N \s_0 = R I^2 \t_0 / k_B T = 10^5$. Thus the final result is
\beq
P[\text{negative fluctuation}] \sim e^{-10^{12}} \ ,
\eeq
which is clearly too small to be observed in an experiment.

Nevertheless, Eq.~(\ref{pnegstima}) suggests some possible strategies to enhance negative
fluctuations: {\it 1)} reduce the observation time; {\it 2)} consider a smaller
system; {\it 3)} reduce $\s_0$. Unfortunately, {\it 1)} is limited by the fact that the
fluctuation relation holds only for large $\t$, \ie $\t \gg \t_0$; so we can
reduce $\t$, but at best we can use $\t \sim 100 \t_0$ if we do not want to observe
finite $\t$ effects \cite{ZRA04a,GZG05}. Note also that in experiments $\t$ is strongly constrained
by the acquisition bandwidth. In any case, given that in our example $N\s_0 \sim 10^5$, even
with $\t/\t_0 = 10^2$ we will not gain much.
Reducing the system size is a very good idea (but might be difficult in experiments):
for this reason numerical simulations to test the fluctuation relation are usually
done for systems of $N < 50$ particles. However, in some cases we might be interested in large
systems if we want to test the chaotic hypothesis, as the deviation from Anosov
behavior might be more evident in small systems.
Finally, we might try to reduce the entropy production in the system. This can be done
by reducing the strength of the applied field (the electric field in our example).
There is a problem however: in the limit of small dissipation, the fluctuation
relation reduces to the Green-Kubo relations that are well known from linear response
theory \cite{Ku57,EM90}. Therefore our experimental test of the fluctuation relation will reduce to
a trivial test of linear response theory, up to $O(E^2)$ corrections.
We will discuss the relation between fluctuation relation and Green-Kubo relations
in the following section.

\subsection{Gaussian fluctuations and the Green-Kubo relation}

The fact that the fluctuation relation reduces to Green-Kubo relations in the
small $E$ limit is disturbing for the purpose of testing the former, but is an 
important check of the consistency of the theory. 
Indeed, a good statistical theory of nonequilibrium stationary states should 
reduce to linear response theory in the limit of small dissipation.
The relation between the fluctuation relation and Green-Kubo relations
has been discovered in \cite{Ga96a,Ga96b}: in the following we will closely
follow the original derivation of \cite{Ga96a}.

\subsubsection{The large deviation function of $\s$ close to equilibrium}

Close to equilibrium the istantaneous entropy production rate has the form 
$\s(t) = E {\cal J}(t)$ \cite{EM90,DGM84}, as discussed above, where $E$ is the driving force
and ${\cal J} = J/(k_B T)$ is the conjugated flux, see Eq.~(\ref{entropyav}).
To compute the function $\z_\io(p)$ it is easier to start from its Legendre
transform $z(\l)$, defined by
\beq 
\label{phidef} 
z_\io(\lambda) = - \lim_{\t \rightarrow \infty} \t^{-1}  
\log \langle \exp[-\lambda \t\sigma_\t ] \rangle =
-\max_p [ \z_\io(p)-\l p \s_+] \ .
\eeq 
The function $z_\io(\l)$ generates the connected moments of $\s_\t$; using
Eq.~(\ref{entropyav}), in the limit $\t\to\io$,
these have the form\footnote{It is convenient for this computation to change convention and
define $\s_\t= \t^{-1} \int_{-\t/2}^{\t/2} dt \, \s(t)$; one can check that
due to time translation invariance the results are unchanged by the convention used.}
\beq\begin{split}
z^{(k)}_\io & \equiv \left. \frac{d^k z_\io}{d\l^k} \right|_{\l=0} =(-1)^{k-1} \lim_{\t \to \io} \t^{k-1} \la \s_\t^k \ra_c  \\ &=
(-1)^{k-1} E^k \lim_{\t \to \io}\t^{-1} \int_{-\t/2}^{\t/2} dt_1 \cdots \int_{-\t/2}^{\t/2} dt_k 
\la {\cal J}(t_1) \cdots {\cal J}(t_k) \ra_c \ .
\end{split}\eeq
The connected correlations $\la {\cal J}(t_1) \cdots {\cal J}(t_k) \ra_c$ are translationally
invariant due to the stationarity of the system and {\it decay exponentially} in the
differences $|t_i - t_j|$ due to chaoticity.

From stationarity it follows that 
$z^{(1)}_\io = E \la {\cal J} \ra$ and as $\la{\cal J}\ra = 0$ in equilibrium, one has
$\la {\cal J} \ra \sim E$ and $z^{(1)}_\io = \s_+ \sim E^2$.
Using stationarity and the exponential decay of the connected correlations one has,
for $k>1$,
\beq\label{moments}
\begin{split}
\lim_{\t \to \io}\t^{-1} \int_{-\t/2}^{\t/2} dt_1 & \cdots \int_{-\t/2}^{\t/2} dt_k 
\la {\cal J}(t_1) \cdots {\cal J}(t_k) \ra_c = \\
=&\int_{-\io}^\io dt_1 \cdots \int_{-\io}^\io dt_{k-1} \la {\cal J}(0) {\cal J}(t_1) \cdots {\cal J}(t_{k-1}) \ra_c 
\equiv {\cal J}^{(k)}_\io \ ,
\end{split}\eeq
and the ${\cal J}^{(k)}_\io$ are finite for $E\to 0$. This means that the $z_\io^{(k)} \sim E^k$
for $k>1$ and 
\beq
z_\io(\l) = z_\io^{(1)} \l + \frac{z_\io^{(2)}}2 \l^2 + O(E^3 \l^3) \ .
\eeq
Using this result and the relation $\z_\io(p) = \min_\l [ \l p \s_+ - z_\io(\l)]$ one can prove
that\footnote{Note that $z^{(2)}_\io$ is negative.}
\beq\label{zsmallE}
\z_\io(p) = \frac{\s_+^2}{2 z_\io^{(2)}} (p-1)^2 - 
\frac{z_\io^{(3)} \s_+^3}{6 \big(z_\io^{(2)}\big)^3} (p-1)^3 + \ldots =
 \frac{\s_+^2}{2 z_\io^{(2)}} (p-1)^2 + O(E^3 (p-1)^3) \ ,
\eeq
\ie that $\z_\io(p)$ is approximated by a
Gaussian up to $|p-1| \sim 1/E$, \ie in an interval whose size grows for
$E\to 0$ \cite{Ga96a,Ga96b}. Note that from Eq.~(\ref{zsmallE})
we also have $\z_\io(0) \sim  \frac{\s_+^2}{2 z_\io^{(2)}} \sim E^2 \sim \s_+$
as anticipated.

\subsubsection{The Green-Kubo relation}

If the function $\z_\io(p)$ is given by the first term in (\ref{zsmallE}), 
the fluctuation relation immediately gives
\beq\label{FRgauss}
\s_+ = -\frac12
z^{(2)}_\io 
\hspace{10pt} \Rightarrow \hspace{10pt} 
\s_+ = \langle \s(t) \rangle = 
\frac12 E^2 \int_{-\io}^\io dt \ \langle {\cal J}(t) {\cal J}(0) \rangle_c \ ,
\eeq  
using (\ref{moments}) for $k=2$.
Recalling that ${\cal J}(t)=J(t)/T$, $\s(t) = E{\cal J}(t)$ and 
$ \langle {\cal J}(t) {\cal J}(0) \rangle_c$ is even\footnote{\label{notainv}This simply follows from
time translation invariance, $ \langle {\cal J}(-t) {\cal J}(0) \rangle_c=
\langle {\cal J}(0) {\cal J}(-t) \rangle_c=\langle {\cal J}(t) {\cal J}(0) \rangle_c$.}
in $t$ one obtains 
\beq 
\langle J \rangle = 
\frac{E}{T} \int_0^\io dt \  \langle J(t) J(0) \rangle \ , 
\eeq 
and to the lowest order in $E$
\beq 
\langle J \rangle_E = 
\frac{E}{T} \int_0^\io dt \  \langle J(t) J(0) \rangle_{E=0} + O(E^2) \ , 
\eeq 
that is exactly the Green-Kubo relation~\cite{Ku57}.

\subsubsection{Extension to many forces: Onsager reciprocity}

If many forces are present, we will have close to equilibrium 
$\s(t) = \sum_i E_i {\cal J}_i(t)$; the derivation above can be
repeated and from the fluctuation relation $\s_+ = -\frac12 z_\io^{(2)}$, 
using
\beq\label{20}
z_\io^{(2)} = -\t \langle \s_\t^2 \rangle_c =
- \sum_{ij} E_i E_j \int_{-\io}^{\io} dt \langle {\cal J}_i(t) {\cal J}_j(0) \rangle \ ,
\eeq
we get
\beq\label{97}
\s_+ = \sum_i E_i \langle {\cal J}_i \rangle = \frac12 
\sum_{ij} E_i E_j \int_{-\io}^{\io} dt \langle {\cal J}_i(t) {\cal J}_j(0) \rangle \ .
\eeq
Defining the transport coefficient $\m_{ij}$ from 
$\langle {\cal J}_i \rangle = \sum_j \mu_{ij} E_j$, and\footnote{\label{onsrep}The equality
$L_{ij}=L_{ji}$ follows from time translation invariance, see footnote \ref{notainv}.}
$L_{ij}= L_{ji} =\frac12 \int_{-\io}^{\io} dt \langle {\cal J}_i(t) {\cal J}_j(0) \rangle$, 
we get from (\ref{97}) the relation
\beq
\frac{\mu_{ij}+\mu_{ji}}2 = L_{ij} \ .
\eeq
However we would like to prove directly the Green-Kubo relation
$\mu_{ij} = L_{ij}$, and Onsager reciprocity $\mu_{ij} = \mu_{ji}$.
This can be done by using the generalized fluctuation relation (\ref{FRestesa})
for the joint distribution of $\s$ and ${\cal J}_i$ \cite{Ga96a,Ga96b}. 
Using Eq.~(\ref{FRestesa}) with $p=\s_\t/\s_+$ and 
\beq
q=\frac1{\t \langle {\cal J}_i \rangle} \int_{-\t/2}^{\t/2} dt {\cal J}_i(t) \ ,
\eeq
and performing a computation similar to the one
in the previous section (see \cite{Ga96a} for the details), 
we get Eq.~(\ref{20}) and the additional relation
\beq
 \langle {\cal J}_i \rangle = \frac12 \sum_j E_j \int_{-\io}^{\io}dt  
\langle {\cal J}_i(t) {\cal J}_j(0) \rangle \ ,
\eeq
which is the Green-Kubo relation for ${\cal J}_i$~\cite{Ku57}. This implies 
$\mu_{ij}=L_{ij}$ and Onsager reciprocity$^{\ref{onsrep}}$ $\mu_{ij} = \mu_{ji}$.

\subsection{Summary and a review of numerical results}

We have seen that even in the case of systems that are guaranteed to verify the
fluctuation relation, a numerical or experimental test might be difficult.
This is due to the fact that:
\begin{itemize}
\item if we wish to verify the fluctuation relation indipendently of linear response
theory, we have to apply a large field, such that the nonlinear terms in (\ref{zsmallE})
are visible for $p \sim 0$: otherwise, if the function $\z_\io(p)$ is Gaussian up to
$p=0$, the fluctuation relation reduces to the Green-Kubo relation;
\item however, if $E$ is large, $|\z(0)|\propto E^2$ is also large, and the probability
to observe negative events becomes very small;
\item the system size must be small, otherwise again $|\z(0)|\propto N$ is large: this can
pose problems if we expect the fluctuation relation to hold only for large enough $N$;
\item the time $\t$ must be large enough for having an interval of $p$ where
the function $\z_\t(p)$ does not depend on $\t$; but if $\t$ is too large, again
negative values of $p$ are less probable and the interval might not contain $p=0$.
\end{itemize}

Despite these difficulties, 
many attempts were made to test the fluctuation relation in numerical simulations of
smooth enough, chaotic and reversible systems which were not guaranteed to strictly satisfy
the requirements of section \ref{sec:ingredients}
\cite{ECM93,ZRA04a,BGG97,BCL98,RM03}. 
Unfortunately in most of these tests the distribution $P(\s)$ was found to be Gaussian.
However these works were very useful for the development of the theoretical concepts
discussed in the previous section, that were largely motivated by the numerical results.

\begin{figure}
\includegraphics[width=8cm]{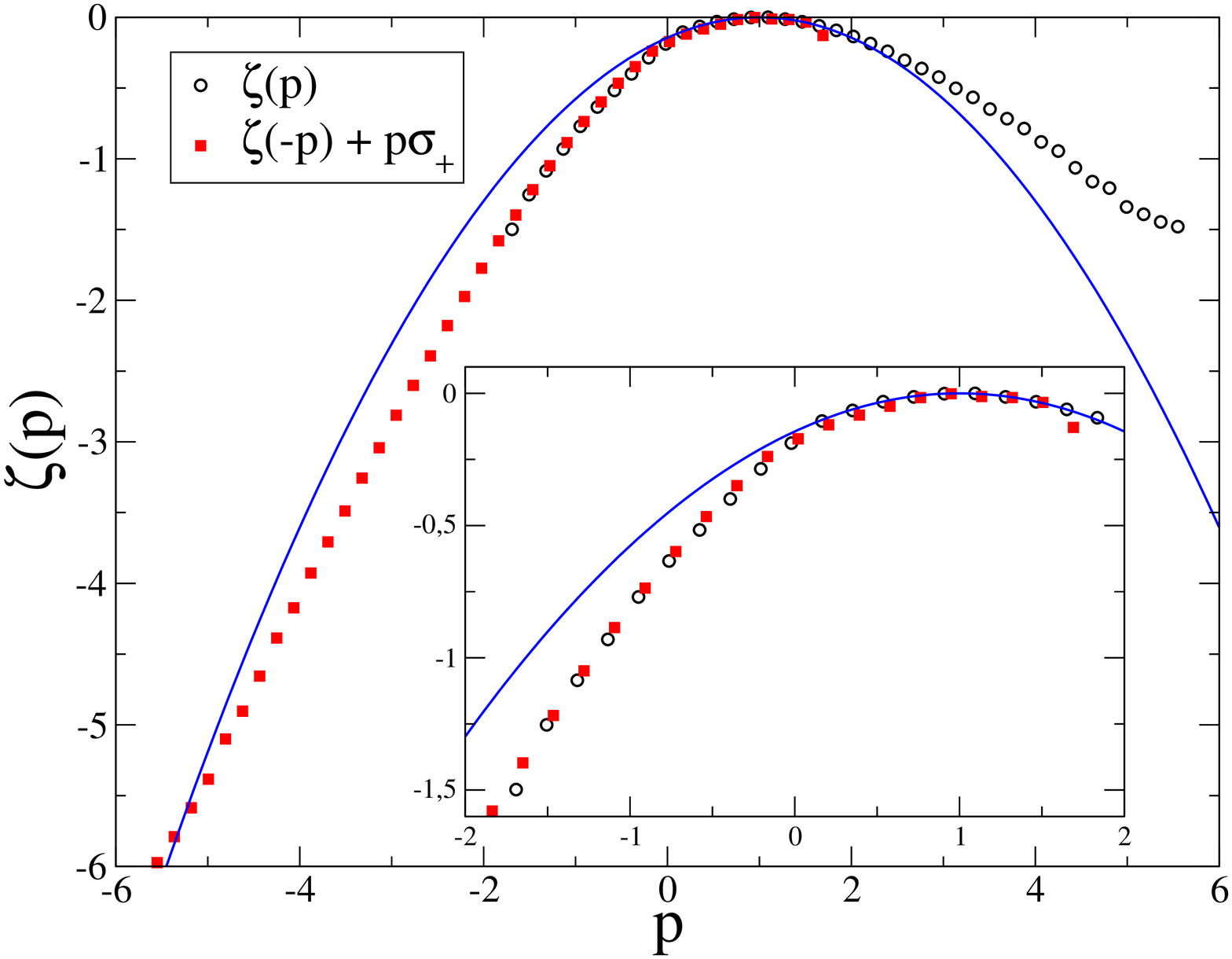}
\includegraphics[width=8cm]{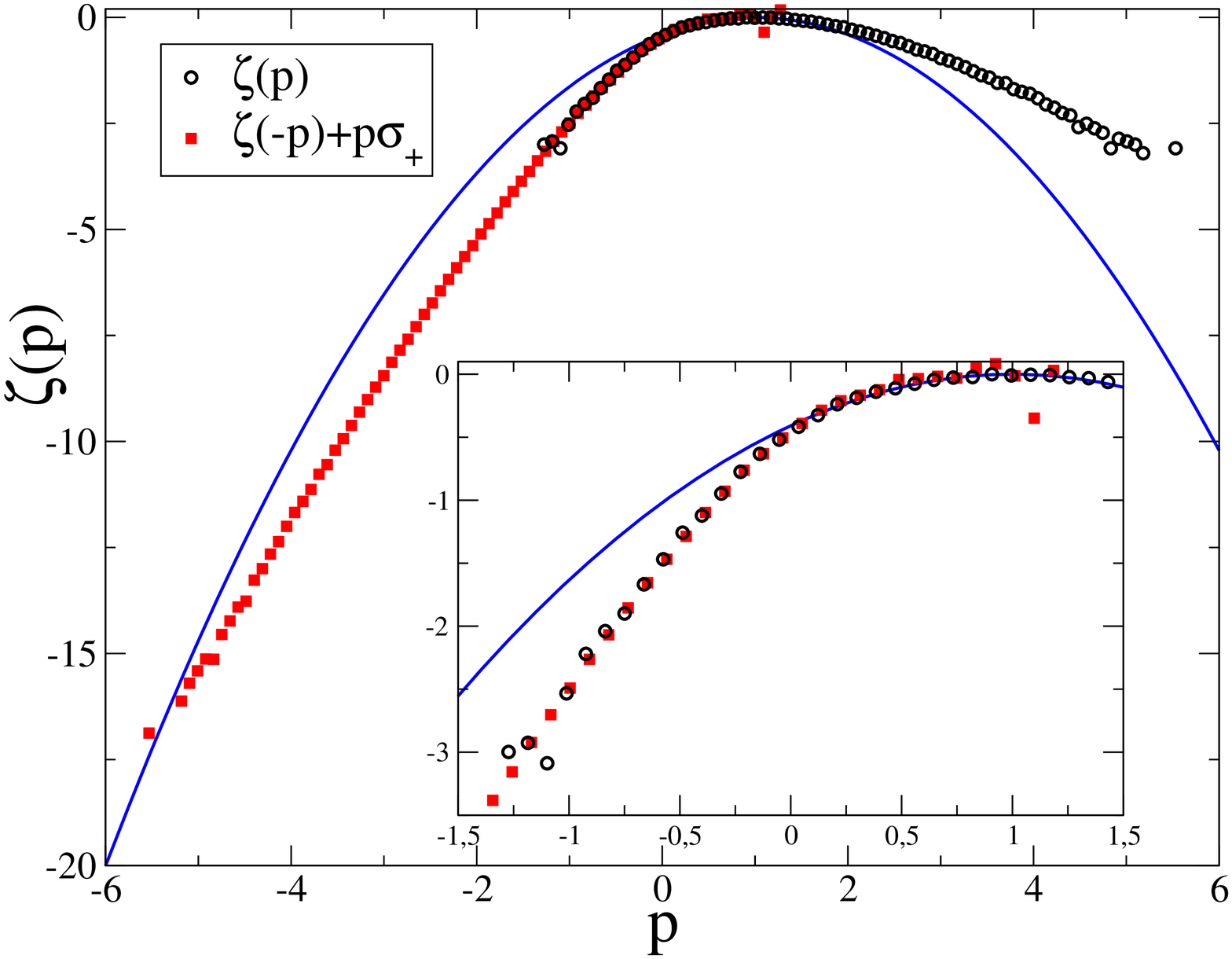}
\caption{Two examples of the function $\z_\io(p)$ measured in a numerical 
simulation~\cite{GZG05}. 
Open circles: $\z_\io(p)$. Filled squares: $\z_\io(-p) + p \s_+$. There are
no free parameters. A Gaussian fit close to $p=1$ is reported as a full line.
Deviations from the Gaussian are seen close to $p=0$. In the inset an enlargment of the region
where circles and square overlap, allowing for a test of the fluctuation relation, is reported.
It was checked \cite{GZG05} that in this interval of $p$ the function $\z_\t(p)$ did not
depend on $\t$, so that it is equal to the true limiting function $\z_\io(p)$.
(Left) $N=8$ Lennard-Jones particles in $d=2$ subject
to a constant force (electric field) with periodic boundary conditions and with
a reversible Gaussian thermostat. 
(Right) Similar system with $N=20$ and $d=3$. See \cite{GZG05} for details on the
simulation.
}\label{figGGZ}
\end{figure}

In \cite{GZG05} it was noted that, using Eq.s~(\ref{zsmallE}) and (\ref{FRgauss}), 
it turns out that the non Gaussian
term in (\ref{zsmallE}) is proportional to ${\cal J}_\io^{(3)} E^3 (p-1)^3$; this suggested
that, keeping fixed $E$ to have a small $\s_+$, one could increase the non Gaussian tails 
of $\z_\io(p)$ by increasing ${\cal J}_\io^{(3)}$, which is related to the nonlinear part
of the transport coefficient. In a fluid of Lennard-Jones like particles,
the nonlinear response is observed to increase on lowering the temperature: this fact was
exploited in \cite{GZG05} where it was possible to verify the fluctuation relation in a 
numerical simulation on a non Gaussian $\z_\io(p)$, see figure~\ref{figGGZ}. 
Note that even if the parameters were
carefully chosen, the finite $\t$ corrections to the function $\z_\io(p)$ had to be taken
into account to extract the correct asymptotic behavior for large $\t$. In particular in the
case of asymmetric distribution the first order correction for large $\t$ is a shift of $p$. 
Thus one can reduce the error by shifting $p$ in such a way that the maximum of $\z_\t(p)$ is
assumed in $p=1$, as we expect in the limit $\t\to\io$. See \cite{GZG05,RM03} for
a detailed discussion.

This problem is specific of systems which admit a natural
equilibrium state and are driven far from equilibrium by some perturbation,
such as a Lennard-Jones fluid. There are however systems (\eg the Navier-Stokes
equations) such that one cannot reach an equilibrium state by tuning some parameter.
For these system there is no obvious linear response theory, and a test of the
fluctuation relation is stringent even if the distribution is found to be Gaussian.
In the case of a reversible version of the Navier-Stokes equations 
(to which the chaotic hypothesis can be applied), 
successful numerical tests of the fluctuation relation 
were performed in~\cite{BPV98,RS99,GRS04}.

The numerical results cited above strongly support the claim that the chaotic hypothesis
can be applied to systems verifying the requirements of section \ref{sec:ingredients}.
Moreover a new very efficient method to sample large deviations has been proposed 
in~\cite{GKP06} and applied to test the fluctuation relation in some systems belonging
to this class, with very promising results.

\section{The effect of singularities: non-smooth systems}
\label{sec:singularities}

Despite these successes, a number of simulations performed using different ensembles
found apparent violations of the fluctuation relation, see \eg \cite{ESR03} and references
therein. These violations where later recognized to be due to the presence of singularities
of the Lennard-Jones potential used in the simulation, that were affecting the measurements
in a subtle way~\cite{Fa02,CV03a,BGGZ06}.
In a system of particles
interacting through a Lennard-Jones potential, the potential energy can be arbitrarily
large if two particles are close enough to each other. If energy is conserved, this
cannot happen, but it can happen, for instance, if only the kinetic energy is kept constant.
This is why violations were observed only when using the isokinetic thermostat.

Consider the power injected into the system by the external forcing, $W(t)$, and the heat
per unit time, $Q(t)$, dissipated by the thermostat. 
Energy conservation implies $Q(t) = W(t) - \dot E(t)$, where $E=K + V$ is the total energy.
Thus if the latter is conserved, $Q=W$, while if kinetic
energy is constant, $Q(t) = W(t) - \dot V(t)$, where $V(t)$ is the
potential energy in the system. The entropy production rate, $\s(t)$, can be defined as
$\s_W(t) = W(t)/T$ or as $\s_Q(t) =Q(t)/T$; there is no {\it a priori} reason for choosing one of the two
definitions. The difference is a total derivative that has
zero average in stationary state, so the average $\s_+$ is unaffected by the choice.

If one considers a microscopic model of the system, it turns out that the
phase space contraction rate, as computed from the equations of motion, 
can be given by any of the two definitions above, depending on which metric one
uses to measure distances in phase space. In fact, the phase space contraction
rate is defined with respect to a given metric $\m(x) dx$ and if one switches
from $\m(x)$ to $\m(x) e^{-\f(x)}$ the phase space contraction rate is changed by
$\dot \f$, see Appendix~\ref{app:defi}. 
For instance, in the case discussed above, the phase space contraction
rate is given by $W(t)/T$ if one considers the contraction of the measure 
$\m(p,q) dp dq = e^{-\b V(q)} \d(K(p) - 3 N T/2) dp dq$, $(p,q)$ being the momenta 
and positions of the particles, and by $Q(t)/T$ if one considers the contraction of 
the volume measure $dp dq$.

What is the effect of the total derivative $\dot V$? If we consider the integrated
variables,
\beq\label{sigmaQW}
\s_{Q\t} = \frac1\t \int_0^\t \frac{Q(t)}T = \s_{W\t} + \frac{V(0)-V(\t)}{\t T} \ .
\eeq
If $V(q)$ is a bounded function, $|V(q)| \leq B$, the difference vanishes as $1/\t$
for large $\t$. This happens uniformly on phase space and it follows that the large
deviation functions of $\s_W$ and $\s_Q$ are equal. Thus, for what concerns asymptotic
statements such as the fluctuation relation, the two definitions are equivalent.
The fluctuation relation does not depend on the measure one chooses
to compute the phase space contraction rate, as anticipated in section~\ref{sec:introduction},
as long as $\s$ is smooth and bounded.

\subsection{The effect of an unbounded total derivative}
\label{sec:zQW}

Let us now see what happens if the function $V$ is not bounded.
The effect of an unbounded total derivative term was first discussed in
\cite{Fa02} and \cite{CV03,CV03a}. Extensions were discussed in 
\cite{BGGZ06,VPBTvW05b,PRV06,Vi06}. Here we will follow and extend
the derivation in \cite{BGGZ06}. As part of the computation is unpublished,
all the details are given in Appendix~\ref{app:saddle}.

We consider the normalized variables $p_Q = \s_{Q\t}/\s_+$ and $p_W = \s_{W\t}/\s_+$,
verifying the relation
\beq
p_Q = p_W + \frac{v_i - v_f}{\t} \ ,
\eeq
having defined the function $v(t) = V(t) / (T \s_+)$. The potential energy is usually
bounded from below, $V(t) \geq B$. As only differences of $V$ matter, we assume that
$B = 0$ by a shift of $V$, without loss of generality.
We assume that $p_W$ has a well defined large deviation function,
\beq
P_\t(p_W) \sim e^{\t \z_W(p_W)} \ ,
\eeq
and we define
\beq
P(v) = e^{-f(v)} \ ,
\eeq
the probability distribution of both $v_i$, $v_f \geq 0$.

\begin{figure}
\includegraphics[width=8cm]{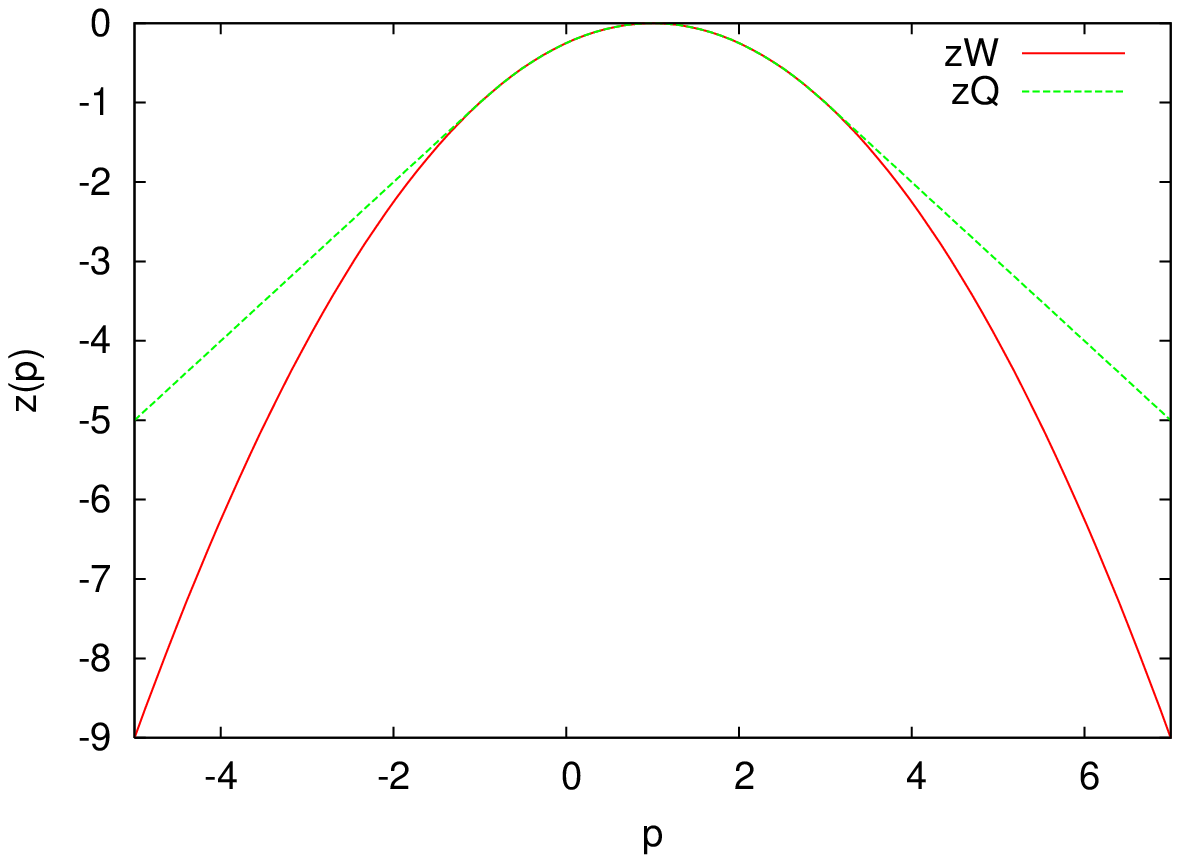}
\includegraphics[width=8cm]{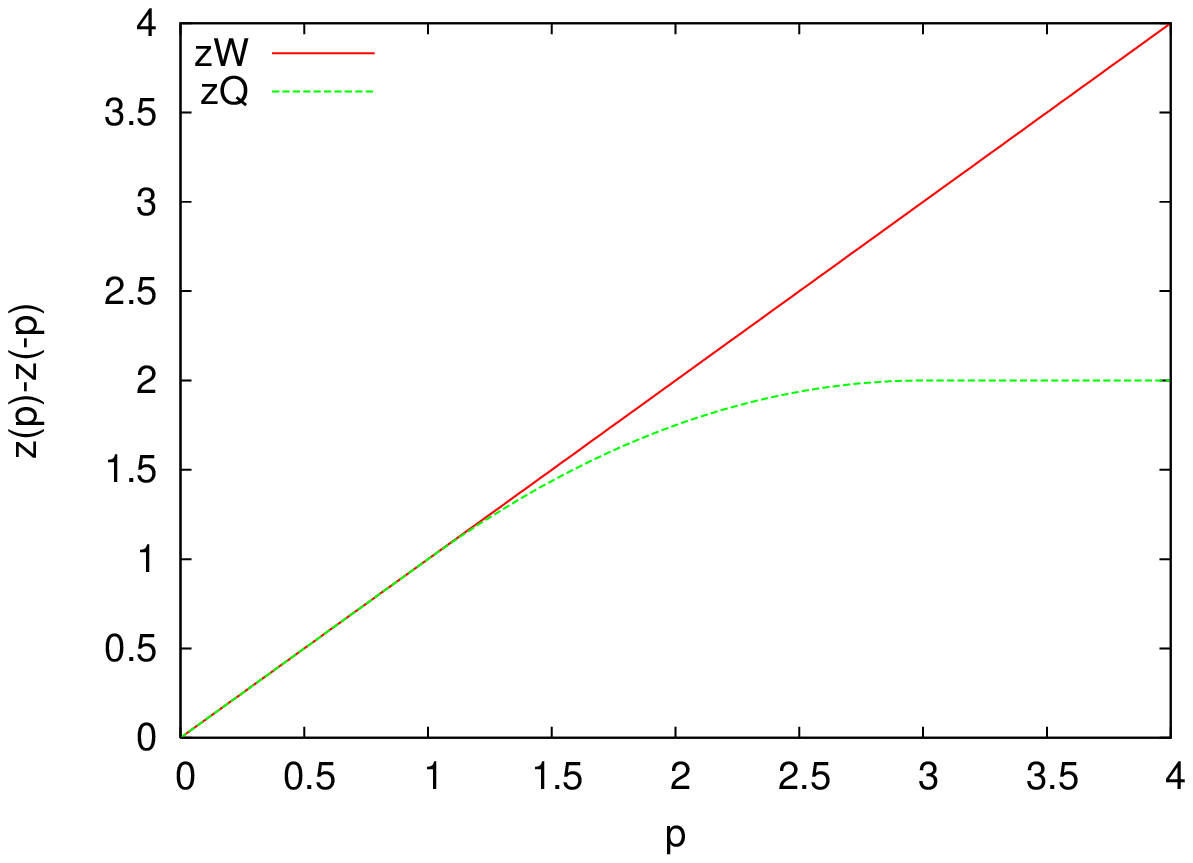}
\caption{The case of exponential tails. In this case $\z_W(p) = -\frac{(p-1)^2}4$ 
verifies the fluctuation relation for $\s_+ = 1$, and $A=1$, hence $P(v)=e^{-v}$. 
The corresponding function
$\z_Q(p)$ coincides with $\z_W(p)$ only for $p\in[-1,3]$, and does not
verify the fluctuation relation for $p > 1$ \cite{CV03a,GC05}.
}
\label{fig:codeexp}
\end{figure}

We assume that $p_W$, $v_i$ and $v_f$ are independent; this follows from the
chaotic hypothesis \cite{BGGZ06}, because for large $\t$, $v_i \propto V(0)$ and $v_f\propto V(\t)$ 
must be uncorrelated, and $p_W$ is an integral of a function evaluated for $t \in [0,\t]$ which
is independent of $V(0),V(\t)$ for most times $t$.
We have then
\beq\label{Pstart}
\begin{split}
P_\t(p_Q) &= \int_{-\io}^\io dp_W \int_{0}^\io dv_i dv_f e^{\t \z_W(p_W) - f(v_i) -f(v_f)}
\d\left( p_Q - p_W - \frac{v_i - v_f}{\t} \right) = \\
&= \t^2 \int_{0}^\io dv_i dv_f e^{\t \z_W(p_Q  - v_i + v_f  ) - f(\t v_i) -f(\t v_f)} \ ,
\end{split}\eeq
where we performed the integration over $p_W$ using the delta function and changed variable
from $v_{i,f}$ to $v_{i,f}/\t$. For large $\t$ we evaluate the integral at the saddle point
and we obtain
\beq\label{max}
\z_Q(p_Q) = \frac1\t \log P_\t(p_Q) \sim \max_{v_i,v_f} \left\{
\z_W(p_Q  - v_i + v_f  ) - \frac1\t [ f(\t v_i) + f(\t v_f)] \right\} \ .
\eeq
The solution of the saddle point equations, detailed in Appendix~\ref{app:saddle}, 
gives the following asymptotic behavior:
\begin{itemize}
\item for superexponential tails, $f(v) \sim A v^\a$, $\a > 1$, one obtains 
$\z_Q(p_Q) = \z_W(p_Q) + O(\t^{-1})$ for all $p_Q$;
\item for exponential tails, $f(v) \sim A v$, we get
\beq\label{zetaexp}
\z_Q(p_Q) = \begin{cases} 
\z_W(p_-) + A(p_Q - p_-) \hskip30pt p < p_- \ , \\
\z_W(p_Q) \hskip90pt p_- < p < p_+ \ , \\
\z_W(p_+) -A(p_Q - p_+) \hskip30pt p > p_+ \ ,
\end{cases}\eeq
where $p_\pm$ are defined by $\z_W'(p_\pm) = \mp A$,
\ie $\z_Q$ coincides with $\z_W$ for $p\in [p_-,p_+]$ and outside this interval 
it is given by its continuation
by straight lines with slope $\pm A$.
\item for subexponential tails,
$f(v) \sim A v^\a$, $0<\a<1$, 
the values $p_\pm$ are defined by
$|p_\pm -1| = \d \sim \t^{-\frac{1-\a}{2-\a}}$:
\beq
\d = \frac{2-\a}{1-\a}\left[ \frac{\z_W''(1)}{ A
    \a(\a-1)}\right]^{\frac1{\a-2}}
\t^{-\frac{1-\a}{2-\a}}
\eeq
and
\beq\label{zetasubexp}
\z_Q(p_Q) =
\begin{cases}
 \z_W(p_Q) \hskip70pt  
|p_Q-1| \leq \d \\
 - A \t^{\a-1} |p_Q-1|^\a \hskip20pt 
|p_Q-1| > \g
\end{cases}
\eeq
for any finite (\ie independent of $\t$) $\g \gg \d$.
In the intermediate regime $\d \leq |p_Q-1| < \g$ the saddle point solution
has to be calculated numerically to interpolate between the two regimes.
The function $\z_Q(p_Q)$ tends to zero for any $p_Q \neq 1$ and coincides with
$\z_W$ only in a small interval around $p_Q=1$, whose amplitude shrinks to $0$
for $\t\to\io$.
\item for power-law tails, $f(v) \sim \b \log v$, we get the same as above with
\beq
\d =| p_\pm -1| = 2 \sqrt{-\frac{\b}{\t \z''_W(1)}} \ .
\eeq
and
\beq
\z_Q(p_Q) \sim - \frac{\b}{\t} \log[ \t |p_Q-1| ]
\eeq
for $|p_Q-1| > \g$ for any finite $\g$.
\end{itemize}
In brief, total derivative terms with superexponential tails are irrelevant, while
exponential tails give a finite modification of $\z_Q$, and subexponential tails
give $\z_Q \to 0$ for all $p_Q$. This makes evident that total derivatives might
have a dramatic effect. 

\begin{figure}
\includegraphics[width=8cm]{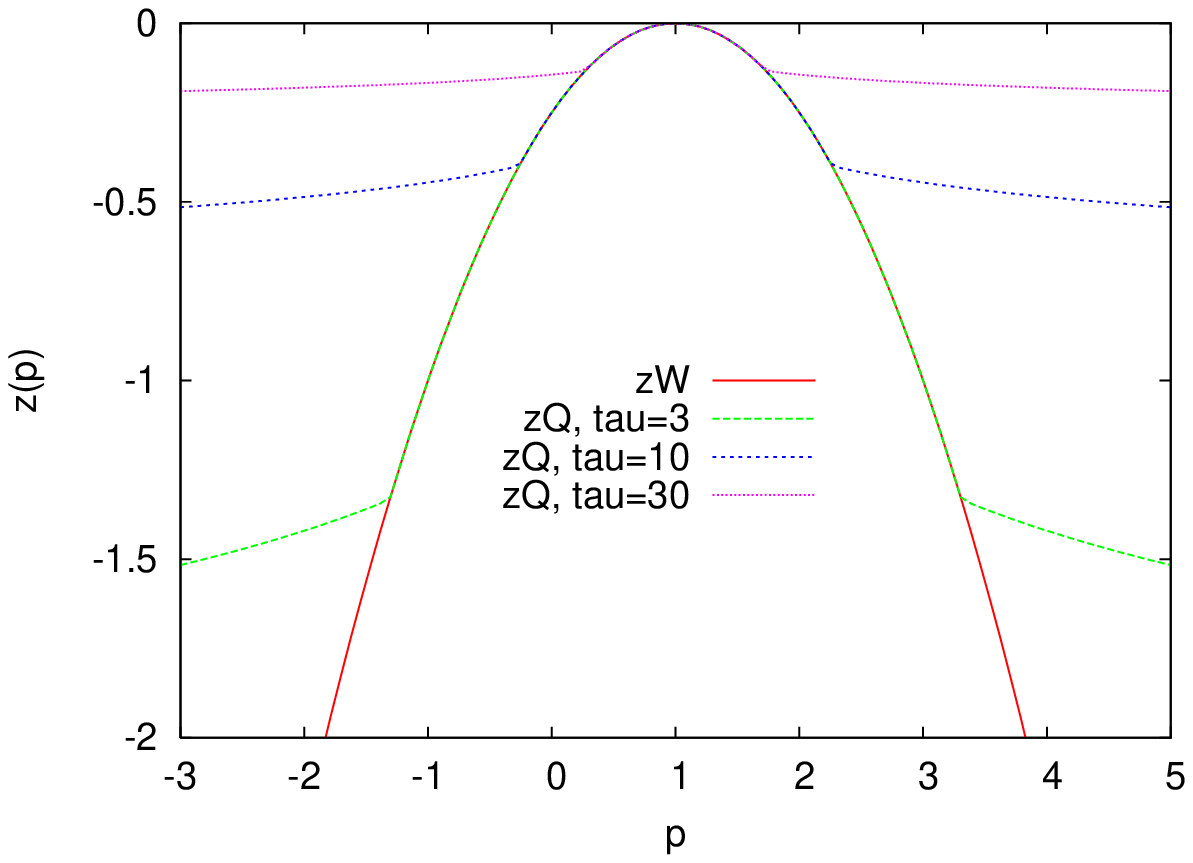}
\includegraphics[width=8cm]{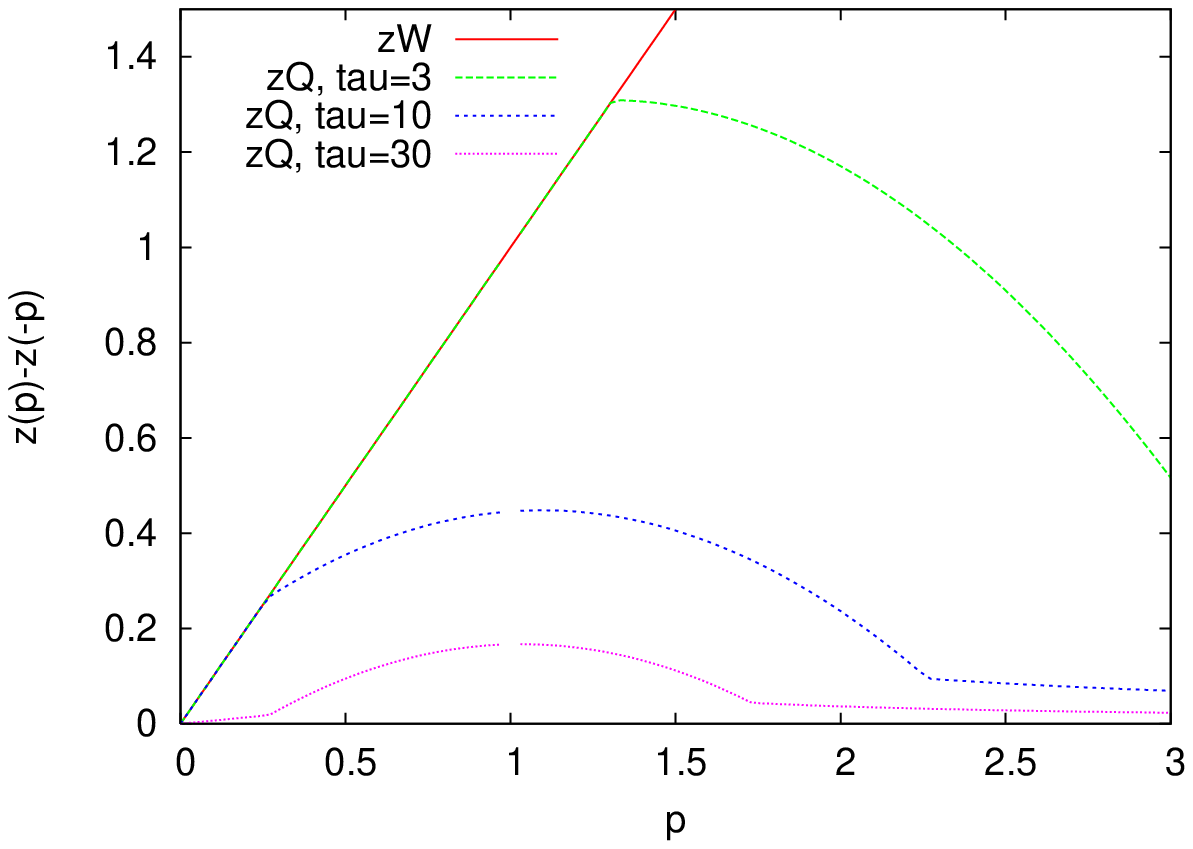}
\caption{The case of power-law tails. $\z_W(p)$ is as in Fig.~\ref{fig:codeexp},
while $f(v) = 2 \log(v)$, \ie $P(v) \sim v^{-2}$. The function $\z_Q(p)$ is
reported for $\t=3,10,30$. In this case for $\t\to\io$, $\z_Q(p) \to 0$. The
fluctuation relation (right panel) is verified only for small $p$ at small $\t$.
For $\t = 30$ one has $p_- > 0$ and the fluctuation relation is never
verified. Note the similarity with the curves in Fig.3 of Ref.~\cite{BCG06}.
}
\label{fig:codepow}
\end{figure}

Assume for instance that in the previous example the function $\z_W$ satisfies the 
fluctuation relation. If $V$ has exponential tails, from Eq.~(\ref{zetaexp}) it follows that
the function $\z_Q$ satisfies the fluctuation relation only for $|p| <\min(p_-,p_+)$, while
for large $p$ the function $\z_Q(p)-\z_Q(-p) \sim$ const, see Fig.~\ref{fig:codeexp} for
an example.
Even worse, in the case of subexponential or power-law tails, $\z_Q$ satisfies the
fluctuation relation only for $|p| < -p_-$, with $p_- \to 1$ for $\t\to\io$. Thus
for large enough $\t$ the fluctuation relation is violated for all $p$, see
Fig.~\ref{fig:codepow}, and 
 $\z_Q(p)-\z_Q(-p) \to 0$. It is interesting to remark that
if $\z_W$ is convex, $\z_Q$ is not if the tails are subexponentials.

The case of exponential tails is particularly relevant because in many cases 
$v \propto V(q)$, the potential energy, whose distribution has tails $e^{-\b V}$,
\ie exponentials. Cases in which the tails are subexponentials might be
also relevant for experiments: for instance the anomalous tails in \cite{BCG06,STX05,FM04}
might be explained by this computation.

\subsection{Removing unbounded total derivatives}

The conclusion of the previous section is that terms that are apparently small
may affect dramatically the large deviations of $\s$ in presence of singularities.
Still the two definitions of entropy production rate seem {\it a priori} equivalent
from a thermodynamic point of view, and have the same average $\s_+$ and the same moments
$\langle \s^k \rangle$ for any finite $k$ (the moments being related to derivatives in $p=1$
of $\z_\io(p)$).

How can we identify {\it a priori} the correct {\it microscopic} definition of $\s$,
satisfying eventually the fluctuation relation? A possible prescription has been discussed
in \cite{BGGZ06} and is based on the proof of the fluctuation relation 
for continuous-time systems given in \cite{Ge98}. The proof is based on a {\it Poincar\'e's
section} mapping the continuous-time flow into a discrete-time map. One considers a set
of {\it timing events}: for instance, in a system of hard spheres, the times at which two
particles collide, or, similarly, in Lennard-Jones systems, the times at which the total
potential energy equals a suitable value $V_0$. In such a way the flow is reduced to a map,
and the proof of the fluctuation theorem for maps \cite{Ga95b} can be applied.
However, the Poincar\'e's section, \ie the definition of timing events, has to be
chosen carefully, in such a way that the resulting map is an Anosov map. This has
been rigorously done for Anosov flows in \cite{Ge98}.

For realistic systems this is clearly very difficult. Anyway, it seems very natural,
if the system has singularities, to choose the timing events in such a way that
singularities are avoided. For instance, if the potential energy $V(t)$ of the system can become
infinite in some points of phase space, one can choose the timing events $t_n$ by $V(t_n) = V_0$.
In the example of the previous section, Eq.~(\ref{sigmaQW}), the two definitions of $\s$
become equivalent on the Poincar\'e's section. More generally, if singularities are avoided
by the Poincar\'e's section, all possible definitions of the phase space contraction rate
$\s(t)$ differ by a bounded total derivative which gives no contribution in the limit $\t\to\io$.

In this way we can remove the ambiguity on the definition of $\s$, at least in the limit
$\t\to\io$. Note that this prescription does not guarantee that the phase space contraction
for the discrete-time system is bounded. The possible values of the latter quantity are
given by
\beq\label{sigmareg}
\s_n = \frac1{t_{n+1}-t_n} \int_{t_{n}}^{t_{n+1}} \s(t) dt \ ,
\eeq
and if $\s(t)$ is not integrable close to a singularity $\s_n$ can be unbounded.
This happens clearly if $\s(t) \propto (t-t_0)^{-\g}$, $\g < 1$, for a singularity
$t_0 \in [t_n,t_{n+1}]$. In most interesting applications, \eg driven Lennard-Jones
systems in contact with a thermostat, it is possible to show that $\s_W$ is bounded
on all phase space, so that the singularities come only from the total derivative and
are obviously integrable as in (\ref{sigmaQW}). In these cases the prescription on
the Poincar\'e's section is enough to remove singularities, and if the system verifies
the conditions {\it 1), 3), 4)} of section \ref{sec:ingredients}, \ie it is reversible,
chaotic and has a dense attractor, the fluctuation relation is expected to hold from the
chaotic hypothesis.

\subsection{A proposal for data analysis in presence of singular terms}

To summarize, singularities might have important effects but these can be controlled,
following the computation and the prescription of the previous sections.

First of all we note that the presence of unbounded terms is manifested
by anomalous tails in $\z_\io(p)$, exponentials or possibily subexponentials. It might
also be manifested by the fact that $\z_\io(p)$ is not convex. A fit to the tails
of the measured $\z_\io(p)$ allows to guess the behavior of the tails of the singular
term, as the behavior of $\z_\io(p)$ for large $p$ is the same as the behavior of
$f(v)$ for large $v$.

Assume that we
have performed an experiment or a numerical simulation whose output is a very long
time trace $\s(t)$, $t\in [0,T]$, where one has chosen a definition of $\s(t)$ in terms of
work, heat, etc. The usual procedure to verify the fluctuation relation
is to fix a time delay $\t$, and integrate $\s(t)$ over subsequent segments of the trajectory
to obtain values of $\s_\t = \t^{-1} \int_{t_0}^{t_0+\t} \s(t) dt$. One then constructs the histogram
$P(\s_\t)$ and the function $\z_\io(p)$ which is used to test the fluctuation relation.

In presence of singularities, the procedure must be modified as follows:
\begin{itemize}
\item One defines in a suitable way\footnote{One should take care here 
in order to avoid introducing biases in the sampling of $\s_k$ (I thank A.~Puglisi
for this remark).
For a system of particles one can for instance define as timing events the instants where
a (suitably defined) ``collision'' takes place.
If one is able to identify the source of singularities, \ie what is the term in the
phase space contraction rate that gives unbounded contributions to $\s$, better Poincar\'e's
sections can be eventually constructed.}
a set of timing events on the trajectory,
$t_n$, such that $\s(t)$ is not singular for $t=t_n$, and that the average
difference between two subsequent timing events is finite, 
$\t_0 = \langle t_{n+1}-t_n \rangle < \io$.
\item One fixes a delay $k$ and constructs values of 
\beq
\s_k = \frac1{t_{n+k}-t_n} \int_{t_n}^{t_{n+k}} dt \s(t) \ .
\eeq
\item Then one constructs the histogram $P_k(\s_k)$ and the large deviation function
$\z_k(p) = k^{-1} \log P_k(p \s_+)$, where $\s_+ = \langle \s_k \rangle$; 
the latter can be used to test the fluctuation
relation in the limit $k\to\io$.
\end{itemize}
If the system is reversible, chaotic and has a dense attractor, the function $\z_k(p)$ should
have a finite limit that will be convex and verify the fluctuation relation (\ref{FR}) 
for large $k$, according to the chaotic hypothesis.

Then, to check the consistency of the analysis, one can recompute the original large deviation
function $\z_\io(p)$. Indeed a guess on the tails of the function $v_{i,f}$ can be done by
looking at the tails of the function $\z_\io(p)$, because as we discussed above the two functions
have the same tails. Using this guess, one can apply the formulas of section \ref{sec:zQW} and
check the self-consistency of the analysis.

The analysis of this section has been confirmed in some cases by numerical simulations 
\cite{ZRA04a,GZG05,Gi06}, and experimentally \cite{GC05}, and seems promising
to interpret recent experiments on turbulent systems \cite{BCG06,STX05}.

Unfortunately, if $\s$ is not integrable between two timing events, \ie (\ref{sigmareg}) is
not bounded, 
at present no general statement can be made. However this seems to be a very 
pathological case which should not be realized in most physical examples.

\section{On the convergence of the transient fluctuation relation to the stationary state
fluctuation relation}
\label{sec:time_scales}

As we discussed in the introduction, even if it may seem that the transient fluctuation
relation implies the fluctuation relation for stationary state in the limit $\t\to\io$,
this is not the case in general. The transient fluctuation relation holds in greater
generality than the stationary state fluctuation relation.
In this section we will discuss some examples that will highlight the importance
of the properties of {\it chaoticity} and {\it transitivity} (\ie the property that
the attractor is dense in phase space) 
to ensure the validity of the fluctuation relation.

As we said
in the introduction, chaoticity is important because it guarantees that initial data
sampled with respect to the volume measure produce trajectories converging
fast enough to the stationary state. 
Indeed, in experiments the system is often prepared at equilibrium, then the driving force
is turned on and the system is let evolve toward the stationary state.
If convergence to the stationary state
is not fast enough, the observed trajectories might not be representative
of the real stationary state, and confusing results may be obtained.

\subsection{Examples of systems such that $P_{eq}$ does not converge to the stationary
state distribution}

\subsubsection{A simple example}
\label{sec:simpleexample}

A very simple example in
which the distribution $P_{eq}$ does not converge to the stationary state distribution 
has been given in~\cite{GC99}. In this
example the transient fluctuation relation holds for any finite time $\t$, including the
$\t\to\io$ limit, for $P_{eq}$, but the real stationary state distribution trivially 
violates the fluctuation relation.

The model describes a free particle evolving in two dimensions
under the action of a constant force $\vec E$ and of a thermostatting
force keeping its kinetic energy constant. If $\vec p$ is the momentum of
the particle, its mass is $1$ and we fix $|\vec p|^2=1$, 
the equation of motion are
\beq
\dot{\vec p} = \vec E - (\vec E \cdot\vec p) \vec p \ ,
\eeq
and introducing the angle $\th$ by $\vec p\cdot\vec E = E \cos \th$, where $E=|\vec E|$,
we can write the equation for $\th$:
\beq\label{eqtheta}
\dot \th = -E \sin \th \ .
\eeq
The dissipated power is given by $W(t)=\vec E \cdot \vec p(t) = E \cos \th(t)$ and is
equal to the entropy production rate, the ``temperature'' $|\vec p|^2$ being equal to $1$.
Note that $W(t) = d\dot\th/d\th$ is the phase space contraction rate for Eq.~(\ref{eqtheta}).

The equation of motion (\ref{eqtheta}) with initial datum $\th_0$ is easily solved,
\beq
\tan \frac{\th(t)}2 = e^{-E t} \tan \frac{\th_0}2 \ .
\eeq
The entropy production rate over the trajectory is given by, defining $s(t) = \tan \th(t)/2$,
\beq\label{AAA}
\begin{split}
\s_\t &= \frac1\t \int_0^\t dt E \cos \th(t) = \frac1\t\int_0^\t dt \frac{\dot\th E \cos \th(t)}{-E \sin \th(t)}  
= - \frac1\t\int_{\th_0}^{\th(t)} \frac{d\th}{\tan \th} \\ 
&= - \frac1\t\log \frac{\sin\th(\t)}{\sin\th_0} = E -  \frac1\t\log\frac{1+s^2_0}{1+e^{-2E\t}s^2_0} \ ,
\end{split}\eeq
and it is easy to see that $\s_\t \in [-E,E]$.

The distribution $P^\th_{eq}(\s_\t)$ is computed imposing uniform distribution over the inital
data $\th_0$. Then, inverting (\ref{AAA}) to express $s_0$ as a function of $\s_\t$, we have
\beq\label{Ptheq}
P^\th_{eq}(\s_\t) = \frac1{2\pi} \left|\frac{d\th_0}{d \s_\t} \right| 
= \left| \frac1{4\pi s_0 (1+s_0^2)} \frac{d s_0}{d \s_\t} \right| =
\frac1{4\pi \sqrt{(e^{\t(E-\s_\t)}-1)(1-e^{-\t(E + \s_\t)})}} \ .
\eeq
It is easy to check that this distribution verifies $P^\th_{eq}(-\s_\t) = e^{-\t\s_\t}P^\th_{eq}(\s_\t)$
for any finite time and that in the limit $\t\to\io$ we have, for $p = \s_\t/E \in (-1,1)$,
\beq
\wt \z_\io(p) = \lim_{\t\to\io} \frac1\t \log P^\th_{eq}(p E) = E \frac{p-1}2 \ ,
\eeq
such that $\wt \z_\io(p)-\wt \z_\io(-p) = E p = \s_+ p $, \ie the limiting distribution
verifies the fluctuation relation. 

The stationary state corresponds to $\th =0$, then $s = 0$ and $\s_\t =E$; thus the
distribution
\beq
P^\th_{st}(\s_\t) = \d(\s_\t - E) 
\eeq
trivially violates the fluctuation relation.

\subsubsection{Another simple example}

In the previous example the stationary state function
$\z_\io(p)$ does not exist because $P^\th_{st}$ is singular. One might argue
that this pathology is responsible for the difference between $\z_\io$ and $\wt\z_\io$.
However this is not the case. Consider as an example\footnote{This example was suggested
by F.Bonetto. A very similar example was discussed in Appendix A1 of \cite{Ga99}.} 
a system whose state variable
is an independent pair $(x,\th)$ with $\th$ evolving according to (\ref{eqtheta}), 
and $x$ describing a reversible Anosov system.
The phase space contraction rate is
\beq\label{120}
\Si(x,\th) = E \cos \th + \s^x(x) \ ,
\eeq
where $\s^x(x)$ is the contraction rate of the Anosov system.
The distribution of $\Si_\t = \t^{-1} \int_0^\t dt[ E\cos\th(t) + \s^x(x(t))]$ is given,
if initial data are sampled according to the volume measure, by
\beq
P_{eq}(\Si_\t) = \int d\s^x_\t P^\th_{eq}(\Si_\t-\s^x_\t) P^x_{eq}(\s^x_\t) \ ,
\eeq
with $P^\th_{eq}$ given by (\ref{Ptheq}). Given that both $P^\th_{eq}$ and
$P^x_{eq}$ verify the transient fluctuation relation (\ref{TFT}), it follows
easily that $P_{eq}(\Si_\t)$ verifies the same relation for any finite $\t$ and
consequently the limit distribution $\wt\z_\io(p) = \lim_{\t\to\io} \t^{-1} \log P_{eq}(p \Si_+)$
verifies the fluctuation relation. Conversely, the distribution in stationary
state is
\beq
P_{st}(\Si_\t) = \int d\s^x_\t P^\th_{st}(\Si_\t-\s^x_\t) P^x_{st}(\s^x_\t) = 
 \int d\s^x_\t \d(\Si_\t-\s^x_\t-E) P^x_{st}(\s^x_\t) =
P^x_{st}(\Si_\t-E) \ ,
\eeq
and, given that $P^x_{st}$ verifies the fluctuation relation, it is easy to see
that the relation is not verified by $P_{st}(\Si_\t)$, neither for finite $\t$ nor
in the limit $\t\to\io$ (this statement can be checked for instance assuming that $P^x_{st}$ is
a Gaussian).

To summarize, in this example the distribution $\t^{-1} \log P_{eq}(\Si_\t)$ converges fast enough
to a limiting distribution $\wt\z_{\io}(p)$ verifying the fluctuation relation. The
true distribution $\z_\io(p) = \lim_{\t\to\io} \t^{-1} \log P_{st}(\Si_\t)$ exists, is analytic,
but is different from $\wt\z_\io(p)$ and in particular does not verify the fluctuation relation.

This example shows that the limiting procedure involved in passing from the transient fluctuation
relation to the stationary state fluctuation relation is very subtle. In this example nothing
seems to go wrong, all the functions are smooth and convergence is fast, but the true stationary
state distribution is different from the limit of the distribution $P_{eq}$. This kind of
subtlety is peculiar to systems that do not display a chaotic behavior at least on some subset 
of the state variables\footnote{Note that the system is chaotic in the sense of having at least
one positive Lyapunov exponent, but the attracting set is not dense in the phase space $(x,\th)$
being concentrated on $\th=0$.}, and might be very difficult to detect in a numerical experiment.

\subsection{Hidden time scales}
\label{sec:time_scale_esempio}

The strange behavior of the examples above can be related to the existence of
a hidden very large time scale\footnote{\label{Kunota}The content of this section is based
on ideas of J.~Kurchan \cite{Ku06}.}.
This time scale can be revealed by adding a small noise term of variance $\e$ to
Eq.~(\ref{eqtheta}). In presence of the noise, the system is able to explore the
full phase space and the fluctuation relation holds in stationary state 
for any finite $\e$.
However, in the limit $\e\to 0$, the time needed is $\t_\e \sim e^{1/\e}$.
For $\t \ll \t_\e$ the fluctuation relation
is violated, while for $\t \gg \t_\e$ it is recovered. Clearly in the limit
$\e\to 0$, $\t_\e \to \io$ and the fluctuation relation is violated for all
$\t$ in stationary state \cite{Ku06}.

As a simple example we consider a two-state system described by a spin
variable $s_t \in \{-1,1\}$. The initial spin is chosen from $P(s_0) \propto e^{h s_0}$, 
then the transition rate $K(s_t,s_{t+1})$ is proportional to $e^{h s_{t+1} -
  \b J(s_t,s_{t+1})}$, where $J(+,-)=1$ and $J=0$ otherwise, \ie 
$J(s,t) = \d_{s,+}\d_{t,-}$.
Then the probability of a
trajectory $s = (s_0,\cdots,s_\t)$ is given by
\beq\label{isingProb}
P[s] \propto e^{h \sum_{t=0}^\t s_t - \b \sum_{t=0}^{\t-1} J(s_t,s_{t+1}) } \ ,
\eeq
\ie the dynamical system\footnote{The reason why we call (\ref{isingProb}) a dynamical
system is that the SRB measure becomes a Gibbs measure for an Ising chain when computed on a Markov
partition. Therefore (\ref{isingProb}) is one of the simplest measures of the form (\ref{SRBrozza}).
Alternatively one can think to this system as a Markov process.}
corresponds to a one-dimensional Ising chain where
only pairs $(+,-)$ give a contribution $1$ to the energy.

Define the time reversed trajectory
$\bar s = (-s_\t,\cdots,-s_0)$; then
\beq\label{isingFR}
P[\bar s] = e^{-2h \sum_{i=0}^\t s_i} P[s] =
e^{-\t \s_\t[s]} P[s] \ ,
\eeq
and this implies the fluctuation relation for the distribution of
\beq
\s_\t[s] = \frac{2h}{\t} \sum_{t=0}^{\t} s_t \ ,
\eeq
for any finite $\b$. 

For infinite $\b$ the transition $(+,-)$ is forbidden:
therefore, in stationary state only the trajectory $s = (+,+,\cdots,+)$ is possible.
Thus $P_{st}(\s_\t) = \d(\s_\t-2h)$ and it does not verify the fluctuation relation, as in the examples
discussed above. Note that transient trajectories of the form $(-,\cdots,-,+,\cdots,+)$ are allowed,
if initial data are extracted according to $P(s_0)\propto e^{hs_0}$,
and the relation (\ref{isingFR}) holds for these transient trajectories. Then the transient 
fluctuation relation holds for $P_{eq}$ also for $\b = \io$ at any finite $\t$, and in the limit $\t\to\io$.
This is exactly the same situation we already discussed in section~\ref{sec:simpleexample}.

The limit of very large but finite $\b$ is particularly interesting: in this limit, the Ising
chain (\ref{isingProb}) develops a large correlation length. 
Consider a segment of trajectory of length $\ell$ beginning with $+$. The segment $(+,\cdots,+)$
has probability $O(1)$. Segments ending in the state $-$ have at least one interface $(+,-)$ which
can be everywhere in $\ell$, then their probability is $\sim \ell e^{-\b}$. Therefore if
one wants to observe a jump $(+,-)$ the segment of trajectory must have length $\ell \sim e^{\b}$.
This is the time we need to wait if we want to observe a jump to the state $-$ which is needed to
observe the fluctuation relation. We conclude that if $\t \ll e^{\b}$ the fluctuation relation
will be violated, while if $\t \gg e^{\b}$ it will be verified. 

In the limit $\b \to \io$, the time scale diverges and this is why the fluctuation relation
is violated also in the limit $\t\to\io$:
the limits $\b\to\io$ and $\t\to\io$ cannot be exchanged~\cite{Ku06}.

\subsection{Transitive Axiom C attractors}

The system (\ref{120}) is a particular case of a more generic situation in which the 
motion of the system on its attractor can be described by a transitive Anosov system.
If the system is reversible, and there is a unique attractor ${\cal A}_+$ and a unique
repeller ${\cal A}_-$ (\ie an attractor for the time-reversed dynamics), the system
is called an Axiom C system. Details can be found in \cite{Ga99,BG97}.

An extended version of the chaotic hypothesis is that, {\it if the attracting set is not
dense in phase space, the system can be regarded as an Axiom C system, \ie the motion
on the attracting set is described by an Anosov system, at least for the purpose of
computing the physically interesting quantities}. The SRB measure describing the
stationary state will be given by an expression similar to (\ref{SRBrozza}), with the
expansion rates computed on the attractor, times a ``delta function'' enforcing the
constraint that the system is on the attractor.

In such systems, the phase space contraction rate can be written as in (\ref{120}):
\beq
\Si(x) = \s_{\cal A}(x) + \s_{\perp}(x) \ ,
\eeq
where $\s_{\cal A}(x)$ describes the phase space contraction rate on the attractor,
while $\s_{\perp}(x)$ describes the part which is orthogonal to the attractor.
One could then be interested in measuring the fluctuation of $\s_{\cal A}(x)$
to test the extended chaotic hypothesis, that implies that $P(\s_{\cal A})$ verifies
the fluctuation relation.
Unfortunately, in general the attractor will be a complicated manifold and the
explicit construction of $\s_{\cal A}(x)$ might be impossible.
In particular, while $\Si(x)$ is related to the entropy production rate, 
it is not obvious to relate $\s_{\cal A}(x)$ to an experimentally accessible quantity.

A possible strategy to access $\s_{\cal A}(x)$, that works under some additional hypothesis, 
has been proposed in \cite{Ga99,BG97}. Unfortunately for the moment a numerical
verification of this proposal is not possible due to prohibitive computational costs
\cite{GZG05}, so we will not discuss this matter in more detail here.

\subsection{Summary}

We discussed some examples in which the transient fluctuation relation holds but the
stationary state fluctuation relation is violated. This is due to the fact that (a subset
of the) system is not chaotic, and might be related to a diverging ``hidden'' time scale
whose existence can be revealed by adding a small noise~\cite{Ku06}.

\section{Irreversible systems}

In the previous section we discussed some
possible violation and/or modification of the fluctuation relation for systems
that violate the requirements of smoothness, chaoticity and ergodicity/transitivity.
The last ingredient which is required, as discussed in section \ref{sec:ingredients},
is the requirement of {\it reversibility}.
This requirement is crucial: indeed, we see from
Eq.~(\ref{FRrozza}) that the fluctuation relation compares the probability of trajectories
$\xx(t)$ having positive entropy production rate with the probability of their time
reversed $I\xx(t)$ having negative entropy production rate. If the system is not reversible,
nothing guarantees that the latter exist: the entropy production could be always positive
and obviously Eq.~(\ref{FRrozza}) does not make sense in this case.

\subsection{Reversible and irreversible models}

We will now discuss some examples \cite{Ga99,Ga97,AFMP01}
in which one can construct two models, one reversible and the
other irreversible, that seems to describe the same physical system.
We will see that the result for the large deviation function 
of the {\it global} entropy production rate, $\z_\io(p)$, is
very different. We will then discuss in what sense
the two models might be equivalent.

\subsubsection{Reversible Gaussian thermostat and constant friction thermostat}

A class of reversible models which is believed to describe
nonequilibrium systems are based on {\it Gaussian thermostats}.
Consider a system of particles described by their position and momenta $(p_i,q_i)$
with equations of motion
\beq\label{Gausstherm}
\begin{split}
m\dot q_i &= p_i \ , \\
\dot p_i &= -\partial_{q_i} V(q) + F_i(q) - \a(p,q) p_i \ . \\
\end{split}\eeq
Here $V(q)$ is the potential energy of interaction between particles, and
$F(q)$ represents an external driving force that does not derive from a potential
and injects energy into the system.
The Gaussian multiplier $\a(p,q)$ is defined by the condition that the total
energy $H(p,q) = \sum_i \frac{p^2_i}{2m} + V(q)$, or the kinetic temperature
$T(p) = \frac{1}{Nd} \sum_i \frac{p^2_i}{m}$, are constant. In the former case one has
\beq
\a(p,q) = \frac{\sum_i p_i F_i(q)}{\sum_i p_i^2} \ .
\eeq
Models belonging to this class are often used to model
electric conduction\footnote{In this case one obtains a reversible version
of the Drude model.}, shear flow, heat flow, and so on, see \eg \cite{EM90} for a review.
The equations (\ref{Gausstherm}) are reversible, the time reversal being simply
$I(p(t),q(t)) = (-p(-t),q(-t))$.

The phase space contraction rate for this equation is easily computed and gives
\beq\label{sigmaGauss}
\s(p,q) = \sum_i \frac{d\dot p_i}{dp_i} + \frac{d\dot q_i}{dq_i} =
 Nd \a(p,q) + O(N^{-1}) =  \frac{\sum_i \dot q_i F_i(q)}{(Nd)^{-1} \sum_i p_i^2/m } =
\frac{W(p,q)}{T(q)} \ , 
\eeq
where $W(p,q)$ is the power injected by the external force.
This result is an example of the identification between phase
space contraction rate and entropy production rate we already discussed\footnote{Note 
that if instead of fixing the total energy we fix the kinetic temperature, the
value of $\s$ and $\a$ change by a term proportional to $\frac{1}{T(p)} \frac{dV}{dt}$.
Thus these two ensembles produce equivalent fluctuations of $\s$, in the thermodynamic
limit, only if the potential
$V$ is not singular. In presence of singularities (\eg for a Lennard-Jones potential)
one has to apply the prescription discussed in section \ref{sec:singularities} to remove
the singular term. This has been shown in \cite{ZRA04a}. A discussion of the difficulties
that one has to face to give a mathematical proof of the equivalence is in \cite{Ru00}.}.

The Gaussian multiplier $\a(p,q)$ is a quantity $O(1)$ in the
thermodynamic limit, and we expect its fluctuations to be $O(1/\sqrt{N})$.
Therefore, in the thermodynamic limit, we can replace $\a(p,q)$ by its average 
$\nu = \langle \a(p,q) \rangle$, in the equation of motion (\ref{Gausstherm}):
\beq\label{Gaussnonrev}
\begin{split}
m\dot q_i &= p_i \ , \\
\dot p_i &= -\partial_{q_i} V(q) + F_i(q) - \nu p_i \ . \\
\end{split}\eeq
But if we do this, the new equations are not reversible! Moreover, the phase
space contraction rate is now $\s_\n(p,q) = N d \nu$ and is always positive, so
certainly the fluctuation relation does not hold for this quantity.
We can ask whether the phase space contraction rate of the {\it reversible}
equations, given by (\ref{sigmaGauss}) and identified with the entropy production rate,
still verifies the fluctuation relation if studied using the {\it irreversible} equations
of motion (\ref{Gaussnonrev}). This is not the case. We have, from the definition of $H$,
\beq
\frac{dH}{dt} =  \sum_i \dot q_i F_i(q) - \n \sum_i \frac{p_i^2}m \ .
\eeq
Using this relation, Eq.~(\ref{sigmaGauss}) becomes
\beq
\s(p,q) = Nd \n + \frac{1}{T(p)}\frac{dH}{dt} \ .
\eeq
The first term is always positive. For large $N$, the second term averaged
over a time $\t$ is roughly equal to $\frac{1}{\t} \frac{\D H}{\la T(p) \ra}$ and
will vanish for $\t\to\io$ if $H$ is bounded\footnote{If the function $H$ is not 
bounded, the fluctuations of $H$ will introduce spurious contributions but still
the fluctuation relation will not hold for $\s(p,q)$.}.
Therefore the integral of $\s(p,q)$ in (\ref{sigmaGauss}) is always positive
for large $\t$, and the fluctuation relation cannot hold for this quantity.

We conclude that
the large deviations of $\s(p,q)$ are different in the
two cases, and in the irreversible case do not verify the fluctuation relation.

This is not surprising, since $\s(p,q)$ is a {\it global} quantity, and it is well
known in equilibrium statistical mechanics that global quantities can have very
different behavior if computed in different ensembles\footnote{The same happens in
equilibrium statistical mechanics: for instance the global energy fluctuates in
the canonical ensemble and obviously does not fluctuate in the microcanonical 
ensemble.}. 

Instead, the two equations (\ref{Gausstherm}) and (\ref{Gaussnonrev}) might be
equivalent for the purpose of computing properties of {\it local} observables,
\ie quatities that depend only on the particles which are in a small box inside
the system, in the thermodynamic limit. In this case the equivalence might hold
also for large deviations. We will discuss this point in the following sections.

\subsubsection{Turbulence and the Navier-Stokes equations}

Another interesting example is the case of the Navier-Stokes equations.
These equations are not reversible. However, it has been conjectured in 
\cite{Ga97,Ga02} that the Navier-Stokes equations might be equivalent,
in some situations, to reversible equations. The equivalence is in the
same spirit discussed in the previous section. 

Numerical results supporting
this conjecture have been given in \cite{BPV98,RS99,GRS04}, where the
validity of the fluctuation relation has been numerically 
verified for the reversible equations.

Note that, an argument similar to the one discussed in the previous section
\cite{AFMP01} leads to the same conclusion, that the fluctuation relation
cannot hold for the {\it global} entropy production in 
irreversible Navier-Stokes equations.

\subsubsection{The granular gas}

The case of granular gases is particularly illustrative. A granular gas is
a system of macroscopic particles (typically of radius $R \sim 1$ mm and mass $m \sim 1$ mg) 
in a container of side $L$ interacting via inelastic collisions 
(typically with restitution coefficient 
$r \sim 0.8 \div 0.9$) and having a large kinetic energy 
$\frac12 m \langle v^2 \rangle \gg m g R$, where $g$ is the acceleration due to gravity.
Energy is injected into the system by shaking the box or vibrating one of its sides.
This system behaves like a gas of hard particles but dissipation is present due to
the inelasticity of the collisions.

The first experiment on such a system \cite{FM04} considered a window
of smaller side $\ell$ inside the box and measured the flux $P_\t$ of kinetic energy entering the window
during a time lapse $\t$. The latter seemed to verify a fluctuation relation at least for
small $p = P_\t / \langle P_\t \rangle$. 
Note that the kinetic energy flux can be written as
\beq\label{granbalance}
P_\t = \D E + D_\t \ ,
\eeq
where $\D E = E(\t)-E(0)$ is the variation of kinetic energy inside the window, while
$D_\t$ is the energy dissipated into the window by inelastic collisions during time $\t$.

A very reasonable model for the inelastic collisions between 
particles~\cite{PVBTvW05,VPBTvW05a,VPBTvW05b,AFMP01}
is to assume that the velocity component parallel to the collision axis is rescaled by $r$, 
in such a way that the relative variation of kinetic energy in the collision is $1-r^2 > 0$.
In such a model, the dissipation $D_\t$ is always positive, and phase space always contracts
so that $\s \geq 0$. It was then recognized~\cite{PVBTvW05}
that, as $D_\t \geq 0$ in this model, the fluctuation relation cannot hold for $P_\t$.
Indeed, if $\D E$ were bounded, the large deviation functions of $P_\t$ and $D_\t$ would
coincide implying that the probability of observing a negative fluctuation of $P_\t$ is
zero for large $\t$. However, $\D E$ is not bounded, and its probability distribution
has an exponential tail, see the discussion in \cite{VPBTvW05b}; 
the method described in section \ref{sec:singularities} can be applied and shows that even
if $D_\t \geq 0$, $P_\t$ can have negative fluctuations. This explains why such fluctuations
were observed in \cite{FM04}. The apparent validity of the fluctuation relation observed
in \cite{FM04} is probably explained by the smallness of the interval of $p$ 
that was accessible to the experiment, in such a way that the function 
$\z_\io(p)-\z_\io(-p)$ appeared as linear in $p$.

The conclusion is that the fluctuation relation does not hold for the quantities $D_\t$
(obviously) and $P_\t$ 
(by (\ref{granbalance}) and the results of section \ref{sec:singularities}), at least
if the model for inelastic collisions used in~\cite{PVBTvW05,VPBTvW05a,VPBTvW05b} is 
accepted.

One can easily construct a model of {\it reversible} inelastic collisions,
in which in a collision particles can gain or loose energy, such that on average
energy is lost but the dynamics is reversible\footnote{F.~Bonetto, private communication.}.
Such a model should give similar results for average quantities but will give a different
result for the large deviations of $P$ and $D$, which will now probabily verify the
fluctuation relation.

\subsection{The time scale for reversibility}

Different models, reversible or irreversible,
give different results when
one computes large deviations of global quantities, and in particular the fluctuation relation
holds for the reversible models but does not hold for the irreversible ones.

If we want to investigate the fluctuations of the global entropy production rate, we
have to ask, given a physical system on which we are performing a measurement,
what is the more appropriate mathematical model, between the reversible and the
irreversible ones, to describe its properties?

Let us discuss this problem for the granular gas we modeled above.
It is reasonable that the real system is
described, at an atomic level, by reversible equations of motion (\ie the Newton 
equations for the atoms constituting the macroscopic particles). In a collisions, the atoms
interact in such a way that kinetic energy of the particles is dissipated by
heating the two particles. Morover sound waves can be emitted as the experiment is
performed in air\footnote{Indeed the experiment is very noisy.}. Thus, {\it in principle},
it can happen that two particles, while colliding, absorb a sound
wave and/or cool spontaneously in such a way that the kinetic energy is augmented
by the collision. Clearly, the probability of such a process is {\it very} small:
it is of the order of $\exp (-N)$, where $N \sim 10^{23}$ is the number of atoms
constituting a particle\footnote{An argument supporting this scaling comes from the ideas
of Kurchan~\cite{Ku06} discussed in section \ref{sec:time_scales}. Indeed, fluctuations
of the internal energy of the two particles are $O(N^{-1/2})$. We can consider them
as a small noise of variance $\ee \propto N^{-1}$ acting during the collision; then
in the limit of small noise the time scale is $\t = \t_0' e^{1/\ee} = \t_0 e^{N}$.}. Thus, 
on the experimental time scale $\t \ll \t_0\exp (N)$
one can safely neglect this possibility, the system is well described by the
model in which velocity are rescaled by $r$ at each collision, and the fluctuation
relation does not hold. However, if one could imagine to wait for a time
$\t \gg \t_0\exp (N)$, the fluctuation relation should be observed to hold.
In this case the violation of the fluctuation relation is related to a mechanism
very similar to the one discussed in section \ref{sec:time_scale_esempio},
\ie it is related to an ``hidden'' diverging time scale (the time scale on
which reversibility of the collisions can be observed).

When the time scale needed for observe
reversibility is not larger that the experimentally accessible time scales,
the system should be well described by reversible models.

\subsubsection{A proposal for an experiment on granular gases}

An example of this procedure was discussed in \cite{BGGZ06b}. We considered a two
dimensional granular gas contained in a box; the bottom of the box is vibrated while
the other sides are fixed. This geometry is different from the one considered in
\cite{FM04} where the whole box is vibrated.

In this situation, a temperature profile is established in the system; kinetic
energy is injected at the bottom and starts to flow trough the system towards the
top of the box, being dissipated in the meanwhile by inelastic collisions.
The energy dissipated by collisions is again always positive, and obviously cannot
satisfy the fluctuation relation.

However, one can look to a {\it different} quantity, namely the flow $J_\t$ of energy through
a small portion of the system located between two lines at height $h$, $h + \d$.
This quantity is not always positive and cannot be expressed as a positive quantity plus
a total derivative, as we did for $P_\t$ in (\ref{granbalance}).
We argued that, in a suitable quasi-elastic limit \cite{BGGZ06b}, see also
\cite{AAF06}, the portion of the system between $h$, $h+\d$ can be thought as 
being equivalent to a system of elastic hard particles in contact with two
thermostats at different temperatures $T_+$, $T_-$. This system can be described
by a reversible model \cite{BGGZ06b}. This analysis
predicts that the fluctuation relation should hold for $J_\t$ at least in a suitable
quasi-elastic limit. It should be possible to verify experimentally (or at least
numerically) this prediction.

\subsection{Ensemble equivalence}

As we discussed in the previous section, if one looks at {\it global}
quantities (\eg in a numerical simulation), the result might depend on the
details of the model which is assumed to describe well the system under investigation.
In particular, the choice between a reversible and an irreversible model
should be motivated by a careful analysis of the involved time scales.

However, this is not so natural. In fact, one of the main results
of equilibrium statistical mechanics is that the relation between the interesting
observables (pressure, density, energy, ...) are {\it independent} of the particular
ensemble one chooses to describe the system. And indeed, in a real experiment,
it is very difficult to distinguish between the system under investigation
and the surrounding environment, and to make a detailed model of the interaction
between them. For instance, one would clearly like that the result does not depend
on the details of the particular device one uses to remove heat from the system, and so on.

To avoid such difficulties, one should consider the system under investigation
as a {\it subsystem} of a larger system, including the thermostats, 
and show that the relations between the interesting
observables of the subsystem do not depend on the details of the model one chooses to
describe the whole system. This would also justify the use of phenomenological models like
(\ref{Gausstherm}), (\ref{Gaussnonrev}), which are clearly unphysical as one assumes
that there is a ``viscous'' term acting on each atom of the fluid.

Proving equivalence of ensembles in nonequilibrium is much more difficult than in
equilibrium, because in the former case {\it dynamics} is important in defining
the ensembles: as the SRB measure (\ref{SRBrozza}) explicitely depends on the
dynamics of the system. For this reason no exact results are available, but only
conjectures~\cite{GR97,Ga00,Ga02} and arguments supporting them \cite{Ru00,Ga00b}, 
see in particular \cite{Ga00b} for a detailed discussion.

The equivalence between two different nonequilibrium ensembles can be defined as
follows. We consider as an example the two models (\ref{Gausstherm}), (\ref{Gaussnonrev});
the equations of motion depend on a number of parameters, such as the density of particles,
the interaction potential $V(q)$, the external forcing $F(q)$ etc.; having fixed these
parameters, the model (\ref{Gausstherm}) depends on the value of the energy $E = H(p,q)$,
while (\ref{Gaussnonrev}) depends on the value of $\n$.
The corresponding ensembles $\EE_E$, $\EE_\n$ are the {\it collection} of the SRB
distributions that describe the stationary states of (\ref{Gausstherm}), (\ref{Gaussnonrev})
at different $E$ or $\n$, respectively.

Consider now a volume $V_0$ inside the container $V$ in which the model is defined, and
a set $\OO(V_0)$ of observables $O_{V_0}(p,q)$ that depend only on the positions
and momenta of the particles inside $V_0$.

The equivalence of $\EE_E$, $\EE_\n$ means that {\it in the limit $N,V\to\io$
at fixed $N/V$ it is possible to establish
a one-to-one correspondence between elements of $\EE_E$ and $\EE_\n$ such that
the averages of all observables $O \in \OO(V_0)$ are equal in corresponding elements}.

In this example, the element $\mu_\nu \in \EE_\n$ corresponding to the element
$\mu_E \in \EE_E$ is defined by the condition that
\beq
\nu(E) = \la \a(p,q) \ra_E = \int dpdq \, \mu_E(p,q) \a(p,q) \ ,
\eeq
and conversely the element $\mu_E$ corresponding to a given $\mu_\n$ is defined
by
\beq
E(\n) =  \la H(p,q) \ra_\n = \int dpdq \, \mu_\n(p,q) H(p,q) \ .
\eeq
Equivalence of the two ensembles means that for all {\it local} observables $O_{V_0} \in \OO(V_0)$ 
it holds
\beq\label{equivalenza}
\lim_{N,V\to\io \, , \, \r=N/V} \la O_{V_0}(p,q) \ra_{\n(E)} = 
\lim_{N,V\to\io \, , \, \r=N/V} \la O_{V_0}(p,q) \ra_{E} \ .
\eeq
In the case of the two ensembles defined by (\ref{Gausstherm}), (\ref{Gaussnonrev}),
the equivalence is supported by the concentration argument discussed above, namely
that for $N,V\to\io$ the fluctuations of $\a(p,q)$ in the isoenergetic ensemble
vanish, see \cite{Ru00,Ga00b} for more detailed discussions.

\subsubsection{A local fluctuation theorem}

Given the definitions above, it is natural to try to 
define a {\it local} entropy production rate and prove for
this quantity a {\it local} fluctuation theorem.
In the example (\ref{Gausstherm}), (\ref{Gaussnonrev}), a possible definition
is a local version of (\ref{sigmaGauss}):
\beq
\s_{V_0}(p,q) =  \frac{\sum_{i : q_i \in V_0} \dot q_i F_i(q)}{(Nd)^{-1} 
\sum_{i : q_i \in V_0} p_i^2/m } =
\frac{W_{V_0}(p,q)}{T_{V_0}(q)} \ .
\eeq
Then, the equivalence conjecture (\ref{equivalenza}), applied to the
average of $O_{V_0}(p,q) = e^{\l \s_{V_0}(p,q)}$ (that generates the probability distribution
of $\s_{V_0}$), implies that the large deviations function
$\z_{V_0}(p)$ is the same in the two ensembles, in the thermodynamic
limit. In principle this function can satisfy the fluctuation relation (\ref{FR})
even if one of the two ensembles is irreversible. This would be a very interesting
result.
Unfortunately, from the theoretical point of view the problem is very difficult.
A tentative theory has been discussed in \cite{Ga99b} and a numerical verification
on a simple model of coupled maps has been reported in~\cite{GP99}. 

In numerical
simulations of more realistic models, see \eg \cite{BCL98}, the results
are much more difficult to interpret. As we discussed in section
\ref{sec:GK}, even in the case of a ``perfect'' system (\ie smooth, reversible,
and transitive) the verification of the fluctuation relation is very difficult
as long as the number of particles is bigger than $\sim 20$. Taking the
limit of $N\to\io$ means that one should look at a subsystem of $\sim 10$ particles
in an environment of, say, $\sim 1000$ particles, and this, at present, 
has prohibitive computational costs. Moreover if the volume $V_0$ is too small,
as required by numerical simulations, one has to take into account nontrivial terms
in the entropy production rate, related to the fluxes (of particles, energy, entropy)
across the surface of $V_0$, see \cite{BGGZ06b} for a tentative discussion of
this problem.

On the other hand, looking at local quantities is very natural in experiments,
see \eg \cite{CL98,CGHLPR04,STX05,FM04}. However the interpretation of these
experiments is not clear for the moment (see \cite{Ga00b} for a tentative 
interpretation of \cite{CL98}), especially because the relation between
the measured quantities and the local entropy production is not straightforward. 
The presence of large tails in the measured
distributions suggests also 
that unbounded terms are affecting the measurements; it would be very interesting
to try to apply the analysis described at the end of section \ref{sec:singularities}
to these data.

\subsection{Summary}

The fluctuation relation cannot hold, even locally, if the time scale to observe reversibility
exceeds the experimentally accessible time scale.
Nevertheless, there are system that might
be well described by reversible equations of motion, if this time scale is not too large.
In these cases the fluctuation relation should hold.
Clarifying this issue is clearly of fundamental importance to interpret experiments
on real nonequilibrium systems.

It is worth to note that even if the fluctuation relation does not hold due
to irreversibility, this does not mean that the chaotic hypothesis does not apply.
If the system is chaotic and smooth enough, still the
stationary state should be described by the SRB measure (\ref{SRBrozza}). It would be
very interesting to derive from the chaotic hypothesis, using the measure (\ref{SRBrozza}), 
other relations, independent of
reversibility, that could be tested in experiments.

\section{Conclusions}

To conclude, we will briefly summarize the main points discussed in this paper.
\begin{enumerate}
\item The chaotic hypothesis states that the fluctuation relation will be generically 
verified by models that
are {\it reversible, smooth, chaotic and transitive}.
\item For such models, numerical simulations have confirmed the validity of these
predictions.
\item A test of the fluctuation relation, even in models verifying the hypotheses above,
is made difficult by the necessity of observing negative values of $p$. 
In particular, one should check that the function 
$\z_\t(p) =\frac1\t \log P(p\s_+)$ is independent of $\t$ for $p \in P$. The interval $P$ must
contain the origin.
\item Negative
fluctuations can be enhanced by reducing the system size $N$, the observation time $\t$,
or the strength of the applied field.
The observation time $\t$ cannot be reduced arbitrarily because the fluctuation
relation holds only for $\t \gg \t_0$, $\t_0$ being the characteristic decorrelation time
of the system.  Eventually we can eliminate some finite $\t$ corrections, \eg by shifting $p$
in such a way that the maximum of $\z_\t(p)$ is assumed in $p=1$.
\item If the applied field is so small that the system is close to equilibrium,
and if the distribution turns out to be Gaussian over the whole accessible interval
in $p$, we are not verifying the fluctuation relation, because in this case it
is equivalent to the Green-Kubo relations.
\item It is important to check not only that $\z_\io(p)-\z_\io(-p)\propto p$, but also
that the proportionality constant is $\s_+$. The linearity in $p$ could simply be due
to the smallness of the accessible interval $P$.
\item The presence of singular terms in $\s$ ({\it non-smooth systems}) 
can change dramatically the behavior
of $\z(p)$. These terms manifest in anomalous large tails; in these cases the function
$\z(p)$ might not be convex. If these terms are present, one should remove them
by applying the procedure discussed in section \ref{sec:singularities}.
\item For systems that are not chaotic or not transitive any kind of strange behavior
can (and has) been observed. For this reason a test of the fluctuation relation is
interesting: it supports the validity of the chaotic hypothesis, \ie that the system
is chaotic and transitive (but obviously does not {\it prove} these properties).
\item If one looks to the {\it global} entropy production rate, the fluctuation relation
will not hold for irreversible systems.
\item However, it is possible that different models which are globally reversible
or irreversible, might be equivalent when observed on a {\it local} scale. If this
is the case, a {\it local fluctuation relation} might hold independently of the model
(reversible or irreversible) one chooses to describe the system on large scale.
\item This is very important for the interpretation of experiments. If the results
were found to depend strongly on the details of the model, one would have to take
into account all the details of the system, including the thermostat, etc. 
\end{enumerate}
Hopefully new experiments will be able to clarify the many open problems, in 
particular the last point.

\section*{Acknowledgments}

This paper is motivated by the invitation to the workshop ``Is it possible
to verify experimentally the fluctuation relation?'', at the {\it Institut
Henri Poincar\'e}, November 30-December 1, 2006. I wish to thank the organizers,
in particular T.~Dauxois, for the invitation, and
all the participants, in particular 
A.~Barrat, S.~Fauve, A.~Puglisi, P.~Visco, F.~van~Wijland for comments and
discussions.

The results reviewed here have been obtained in collaboration with
L.~Angelani, F.~Bonetto, L.~Cugliandolo, J.~Kurchan,
G.~Gallavotti, T.~Gilbert, A.~Giuliani, G.~Ruocco.
I am particularly indebted to F.~Bonetto and G.~Gallavotti for many suggestions and for
pointing out errors and imprecisions in early versions of this paper, and to
J.~Kurchan for discussions on the content of section \ref{sec:time_scales}.

%%%%%%%%%%%%%%%%%%%%%%%%%%%%%%%%%%%%%%%%%%%%%%%%%%%%%%%%%%%%%%%%%%%%%%%%%%%%%%%%%%%
%%%%%%%%%%%%%%%%%%%%%%%%%%%%%%%%%%%%%%%%%%%%%%%%%%%%%%%%%%%%%%%%%%%%%%%%%%%%%%%%%%%
%%%%%%%%%%%%%%%%%%%%%%%%%%%%%%%%%%%%%%%%%%%%%%%%%%%%%%%%%%%%%%%%%%%%%%%%%%%%%%%%%%%
%%%%%%%%%%%%%%%%%%%%%%%%%%%%%%%%%%%%%%%%%%%%%%%%%%%%%%%%%%%%%%%%%%%%%%%%%%%%%%%%%%%
%%%%%%%%%%%%%%%%%%%%%%%%%%%%%%%%%%%%%%%%%%%%%%%%%%%%%%%%%%%%%%%%%%%%%%%%%%%%%%%%%%%
%%%%%%%%%%%%%%%%%%%%%%%%%%%%%%%%%%%%%%%%%%%%%%%%%%%%%%%%%%%%%%%%%%%%%%%%%%%%%%%%%%%
\appendix

\addcontentsline{toc}{section}{Appendix}

\section{Lyapunov exponents and the phase space contraction rate}
\label{app:defi}

We considered a system described by the equation of motion $\dot x = F(x)$
where $x = (x_1,\cdots,x_N)$. Consider a segment of trajectory $\xx(t)$,
$t\in[0,\t]$ (time can be continuous or discrete).
The Jacobian matrix of the trasformation $dx(0) \to dx(\t)$ is given by
\beq
\partial S_{ij}[\xx(t)] = \frac{\partial x_i(\t)}{\partial x_j(0)} \ ,
\eeq
has eigenvalues $s_i[\xx(t)]$, $i = 1,\cdots,N$. The Lyapunov exponents
can be defined as
\beq
\l_i[\xx(t)] = \frac1\t \log |s_i[\xx(t)] | \ .
\eeq
Thus positive Lyapunov exponents correspond to expanding directions, while
negative exponents correspond to contracting directions.

The expansion factor is
\beq
\L_u[\xx(t)] = e^{\t \sum_{i,+} \l_i[\xx(t)]} = |\det \partial S_{ij}[\xx(t)]_u | \ ,
\eeq
where the sum $\sum_{i,+}$ is restricted only to positive exponents and
the determinant is restricted to the unstable directions (corresponding to the positive
exponents). The contraction factor is 
\beq
\L_s[\xx(t)] = e^{\t \sum_{i,-} \l_i[\xx(t)]} = |\det \partial S_{ij}[\xx(t)]_s | \ ,
\eeq
so that the total phase space contraction rate is
\beq\label{Adef}
e^{-\t \s[\xx(t)]} = e^{\t \sum_{i} \l_i[\xx(t)]} = |\det \partial S_{ij}[\xx(t)] | 
= \L_u[\xx(t)]\L_s[\xx(t)] \ .
\eeq 

For very small times\footnote{For discrete times clearly the following derivation
is not possible and one has instead $\s[\xx(t)] = -\frac1\t \sum_{t=0}^{\t-1}
\log |\det \partial S(x)|$.}
$\t = dt$ and initial datum $x$ we have
\beq
\partial S_{ij}[x,dt] = \d_{ij} + dt \frac{\partial F_i(x)}{\partial x_j} \ ,
\eeq
so that
\beq
|\det  \partial S_{ij}[\xx(t)] | = 1 +dt \sum_i \frac{\partial F_i(x)}{\partial x_i} +
O(dt^2) \ ,
\eeq
and (\ref{Adef}) becomes
\beq
\s(x) = - \sum_i \frac{\partial F_i(x)}{\partial x_i} = - \nabla \cdot F(x) \ .
\eeq
Then it follows 
\beq
\s[\xx(t)] = - \frac1\t \int_0^\t dt \nabla \cdot F(x(t)) \ .
\eeq

For the time reversed trajectory $\xx(\t-t)$, it is easy to see that
\beq
\partial S_{ij}[\xx(\t-t)] = \frac{\partial x_i(0)}{\partial x_j(\t)} =
\left[\partial S_{ij}[\xx(t)]\right]^{-1}
\ ,
\eeq
and the sign of the Lyapunov exponents are exchanged\footnote{Note that the time
reversal might involve a transformation $U$ such that $U^2 = 1$, \ie the time
reversed trajectory is $I\xx(t) = U\xx(\t-t)$. For instance in the case of an Hamiltonian system
one has to change sign to the momenta.
In this case $\partial S[I\xx] = U \partial S[\xx]^{-1} U$ and if $v_i$ is an eigenvector
of $\partial S[\xx]$ with eigenvalue $s_i$, then $Uv_i$ is an eigenvector of 
$\partial S[I\xx]$ with eigenvalue $s_i^{-1}$.}, as it is needed for the derivation
of (\ref{FRrozza}).

Finally note that $\s(x)$ is defined by $\frac{d}{dt} dx = -\s(x) dx$. Then,
if instead of $dx$ we use the measure $e^{-\f(x)}dx$, it is easy to see that
\beq
\s'(x) = -\frac{1}{e^{-\f(x)}dx}\frac{d}{dt} e^{-\f(x)} dx =  \s(x) + \dot\f \ ,
\eeq
\ie a change of metric changes $\s(x)$ by a total derivative.

\section{The transient fluctuation relation}
\label{app:TFT}

Here we sketch the derivation of Eq.~(\ref{TFT}) for the microcanonical
ensemble. The probability $P_{eq}\{\s[\xx(t)]=\s\}$ is proportional to the
volume $V_0$ of initial data $x_0$ such that the subsequent trajectory gives a
total phase space contraction $\s[\xx(t)]=\s$. Then, by definition of the phase
space contraction rate, the volume of the set of the final points $x_\t=x(\t)$ will
be $V_\t = e^{\t\s} V_0$. Each point\footnote{Or eventually $Ux(\t)$, but as $U^2=1$
the map $U$ conserves volume.}
 $x(\t)$ is the starting point of a
time-reversed trajectory having phase space contraction rate $\s[\xx(t)]=-\s$, so the
volume $V_\t$ is proportional to the probability $P_{eq}\{\s[\xx(t)]=-\s\}$, and
Eq.~(\ref{TFT}) follows.

\section{A precise definition of Anosov systems}
\label{app:anosov}

An Anosov system is defined as follows \cite{GBG04}.
Given a compact, smooth and boundaryless manifold $M$ (phase space), a map
$S \in C^2(M)$ is an {\it Anosov map} if:

\noindent
{\it (1)} For all $x \in M$ the tangent plane to $M$ in $x$, $T_x$, admits a
decomposition $T_x = T^s_x \oplus T^u_x$, such that

\noindent
{\it (2)} the planes $T^{s,u}_x$ vary continuously w.r.t. $x$, \ie the vectors
defining them are continuous functions of $x$;

\noindent
{\it (3)} the angle between $T^s_x$ and $T^u_x$, defined as the minimum angle
between a vector in $T^s_x$ and a vector in $T^u_x$, is not vanishing;

\noindent
{\it (4)} defining $\partial S$ the linearization matrix of $S$ in $x$, \ie
$S(x+\ee v) = S(x) + \ee \ \partial S(x) \cdot v + O(\ee^2)$, $x \in M$, $v\in T_x$,
$\ee \in \RRR$ small, the planes $T_x^{s,u}$ are conserved under $S$, \ie if
$v \in T_x^{s,u}$, then $\partial S(x) \cdot v \in T^{s,u}_{Sx}$;

\noindent
{\it (5)} for all $x \in M$ and for all $v \in T^s_x$ one has 
$|\partial S(x)^k v|_{S^k x} \leq \L^{-k} C |v|_x$, while for
all $v \in T^u_x$ one has 
$|\partial S(x)^{-k} v|_{S^{-k} x} \leq \L^{-k} C |v|_x$, $| \bullet |_x$ being
the norm on $T_x$, for some constants $C,\L > 0$;

\noindent
{\it (6)} there is a point $x$ which has a dense orbit in $M$.

\noindent
The hypotheses above imply that it is possible to identify two families of smooth
manifolds $M^{s,u}$ in $M$, such that $T^{s,u}_x$ are the tangent plane to
$M^{s,u}$ in $x$, and such that points on $M^s$ tend to converge exponentially
while points on $M^u$ tend to diverge exponentially under the action of $S$.

\noindent
This means that for each $x \in M$ there is a stable manifold $M^s_x$ such that
for all $y \in M^s_x$ one has $d(S^k x,S^k y) \leq \L^{-k} C d(x,y)$, and
an unstable manifold $M^u_x$ such that
for all $y \in M^u_x$ one has $d(S^{-k} x,S^{-k} y) \leq \L^{-k} C d(x,y)$, where
$d(x,y)$ is the distance on $M$, see Fig.~\ref{fig:anosov}.

\section{Saddle point equation for the distribution $\z_Q$}
\label{app:saddle}

This computation is reported in full detail as it has not been
previously published.
The saddle point equation corresponding to (\ref{max}) are
\beq\label{SP}\begin{split}
-\z_W'(p_Q  - v_i + v_f  ) &=  f'(\t v_i) \ , \\
\z_W'(p_Q  - v_i + v_f  ) &=  f'(\t v_f) \ .
\end{split}\eeq
For large values of its argument, $f(v)$ must be an increasing function, then
$f'(v) > 0$. We will see that the saddle point values of $v_i$ and $v_f$ are small.
Then, if $p_Q > 1$, the second equation (\ref{SP}) has no solution\footnote{The 
unconvinced reader can check this statement by explicit computation in simple cases.}, 
because $\z_W'(p_Q) < 0$; this means that $v_f$ will stick to the boundary of the
integration region, $v_f=0$. Conversely, for $p_Q < 1$, we have $v_i=0$. Finally,
we get for $p_Q > 1$
\beq\label{maxM1}
\z_Q(p_Q) \sim \max_{v_i} \left\{
\z_W(p_Q  - v_i  ) - \frac1\t f(\t v_i) \right\} \ ,
\hskip30pt
-\z_W'(p_Q  - v_i ) =  f'(\t v_i) \ ,
\eeq
and for $p_Q > 1$
\beq\label{maxm1}
\z_Q(p_Q) \sim \max_{v_f} \left\{
\z_W(p_Q  + v_f  ) - \frac1\t f(\t v_f) \right\} \ ,
\hskip30pt
\z_W'(p_Q  + v_f ) =  f'(\t v_f) \ .
\eeq
In the following we will discuss the behavior for $p_Q >1$, the other case can
be discussed in the same way.

\vskip10pt
\noindent
{\bf Superexponential tails}

First we assume that $f(v) \sim A v^\a$, $\a>1$, for
$v \to \io$, \ie the probability of $v$ decays faster than exponentially.
We have
\beq
-\z_W'(p_Q  - v_i ) =  \a A (\t v_i)^{\a-1} \ .
\eeq
Clearly $v_i$ must be small, and we get
$v_i = \t^{-1} [-\z_W'(p_Q)/\a A]^{1/(\a-1)} \propto \t^{-1}$, and\footnote{Note that in
this case the result is 
inconsistent with the initial hypothesis that $\t v_i$ is large. The computation
has to be repeated by considering the full function $f(v)$; we find
$v_i = \t^{-1} (f')^{-1}[-\z_W'(p_Q)]$ still $\propto \t^{-1}$, thus the result
is unchanged.}
$\z_Q(p_Q) = \z_W(p_Q) + O(\t^{-1})$ for all $p$.

\vskip10pt
\noindent
{\bf Exponential tails}

For $f(v) \sim A v$, we get
\beq
-\z_W'(p_Q  - v_i ) =  A \ ,
\eeq
As $v_i$ must be positive, and $\z_W'(p_Q )$ is negative and decreasing on incresing $p_Q$,
the latter equation has a solution only if $\z_W'(p_Q) < -A$, otherwise $v_i=0$.
Calling $p_+$ the solution of $\z_W'(p_+) = -A$,
we get $v_i^* = p_Q - p_+$ and
\beq
\z_Q(p_Q) = \begin{cases} \z_W(p_Q) \hskip50pt p < p_+ \ , \\
\z_W(p_+) -A(p_Q - p_+) \hskip30pt p > p_+ \ ,
\end{cases}\eeq
\ie $\z_Q$ coincides with $\z_W$ up to $p=p_+$ and then it is given by its continuation
as a straight line with slope $\z'_W(p_+)$. A similar result is obtained for $p_Q <1$.

\vskip10pt
\noindent
{\bf Subexponential tails}

For $f(v) \sim A v^\a$, $0<\a<1$, the saddle point equation becomes
\beq
-\z_W'(p_Q  - v_i ) =  \t^{\a-1} A \a v_i^{\a-1} \ .
\eeq
In addition to the previous solution $v_i \propto \t^{-1}$, we can look for a solution
such that $p_Q - v_i = 1 + \ee$ with $\ee$ small. Indeed as $\z_W'(1) = 0$ the left
hand side will be small, and as $v_i \sim p_Q -1$ is finite 
the left hand side will also be small.
We get then
\beq\label{grap}
-\z_W'(p_Q  - v_i ) \sim -\z_W''(1) (p_Q-1 - v_i) = \t^{\a-1} A  \a v_i^{\a-1} \ ,
\eeq
and we recall that $-\z_W''(1) > 0$. A graphical analysis of the previous equation reveals that
it has no solution as long as $p_Q < p_+$; the value of $p_+$ is defined by the system
\beq
\begin{cases}
-\z_W''(1) (p_Q-1 - v_i) = \t^{\a-1} A  \a v_i^{\a-1} \ , \\
\z_W''(1) = \t^{\a-1} A  \a(\a-1) v_i^{\a-2} \ ,
\end{cases}
\eeq
whose solution is easily found
\beq
\begin{cases}
p_+ -1 = v^*_i \frac{2-\a}{1-\a} \ , \\
v^*_i =\left[ \frac{\z_W''(1)}{\t^{\a-1} A  \a(\a-1)}\right]^{\frac1{\a-2}} \ .
\end{cases}
\eeq
From the above expression we find that $p_+ - 1 \sim v_i^* \sim \t^{-\frac{1-\a}{2-\a}}$.
This is consistent with the hypotheses we made above. Note also that $\t v_i^*$ is large
consistently with the assumption that only the large tail of $f(v)$ is relevant for this
calculation.

%\begin{figure}
%\includegraphics[width=10cm]{grafico.eps}
%\caption{Graphical representation of Eq.(\ref{grap}) for $\z_0''(1)=-1$,
%  $\a=0.5$, $A=1$ and $\t=10$. The saddle point correspoding to the maximum
%is the largest one. For $p<p_+$ there is no solution; the solution appear at
%$p=p_+$ and for large $p$ the largest solution tends to the zero of the left
%hand side, \ie $v_i^* \sim p-1$.
%}
%\end{figure}

For $p_Q < p_+$ the saddle point equation has no solution, it follows that $v_i = 0$ and
$\z_Q(p_Q) = \z_W(p_Q)$. For $p_Q > p_+$, the equation has two solution. As discussed above, 
for finite $p_Q-1$ one solution is $v_i \sim \t^{-1}$ while the other one is $v_i \sim
p_Q-1-\ee$ for small $\ee > 0$. The latter
gives the maximum in (\ref{max}) while the former is a minimum. The value of $\ee$ is given
by
\beq
\ee = \frac{A \a (p_Q-1)^{\a-1}}{-\z_W''(1) \t^{1-\a}} \ .
\eeq
Substituting this value into (\ref{max}) we get
\beq
\z_Q(p_Q) \sim \frac{\z_W''(1)}2 \ee^2 - A \t^{\a-1} (p_Q-1)^\a \sim 
- A \t^{\a-1} (p_Q-1)^\a + O(\t^{2(\a-1)}) \ ,
\eeq
\ie $\z_Q(p_Q)$ is dominated by the tail of $P(v)$ and one gets $\z_Q(p_Q) \to 0$ for all $p_Q$.

For $p <1$, changing $v_i \to v_f$, we get
\beq
\begin{cases}
1-p_- = v^*_f \frac{2-\a}{1-\a} \\
v^*_f =\left[ \frac{\z_W''(1)}{\t^{\a-1} A  \a(\a-1)}\right]^{\frac1{\a-2}}
\end{cases}
\eeq
\ie $1-p_- = p_+ -1 $ so that $p_-$ and $p_+$ are symmetrically distributed around $1$,
and for $p_Q < p_-$
\beq
\z_Q(p_Q) \sim - A \t^{\a-1} (1-p_Q)^\a + O(\t^{2(\a-1)}) \ .
\eeq

To summarize, we obtain two values of $p$, $p_\pm$, defined by
$|p_\pm -1| = \d \sim \t^{-\frac{1-\a}{2-\a}}$:
\beq
\d = \frac{2-\a}{1-\a}\left[ \frac{\z_W''(1)}{ A
    \a(\a-1)}\right]^{\frac1{\a-2}}
\t^{-\frac{1-\a}{2-\a}}
\eeq
such that
\beq
\begin{cases}
\z_Q(p_Q) = \z_W(p_Q) \hskip70pt  
|p_Q-1| \leq \d \\
\z_Q(p_Q) = - A \t^{\a-1} |p_Q-1|^\a \hskip20pt 
|p_Q-1| > \g
\end{cases}
\eeq
for any finite $\g \gg \d$.
In the intermediate regime $\d \leq |p_Q-1| < \g$ the saddle point solution
has to be calculated numerically to interpolate between the two regimes.

\vskip10pt
\noindent
{\bf Power-law tails}

If $v_{i,f}$ have power-law tails for large $v$, $P(v) \sim \frac{1}{v^\b}$, 
we have $f(v) \sim \b \log(v)$,
and the saddle point equations is for $p_Q>1$
\beq
-\z'_W(p_Q - v_i) = \frac{\beta}{\t v_i} \ ,
\eeq
and as before we can expand the l.h.s. around $p_Q-v_i \sim 1$ as the
r.h.s. is small:
\beq\label{sadlog}
-\z''_W(1) (p_Q-v_i-1) = \frac{\beta}{\t v_i} \ .
\eeq
A solution appears for $p_Q>p_+$ where
\beq
\d \equiv p_+-1 = 2 \sqrt{-\frac{\b}{\t \z''_W(1)}} \ .
\eeq
Eq.(\ref{sadlog}) becomes
\beq
v_i^2 - v_i (p_Q-1) + \frac{\d^2}{4} =0
\eeq
with solutions
\beq
v_i = \frac{(p_Q-1)}{2} \pm \sqrt{\frac{(p_Q-1)^2}{4} - \frac{\d^2}{4}} \ .
\eeq
As before the correct solution is the largest one,
\beq
v_i^*(p_Q) = \frac{(p_Q-1)}{2} + \sqrt{\frac{(p_Q-1)^2}{4} - \frac{\d^2}{4}}
\eeq
and as before it behaves as 
\beq
v^*_i(p_Q) \sim p_Q-1- \frac{\d^2}{4 (p_Q-1)} = p_Q-1-\ee
\eeq
for large $\t$ (i.e. small $\d$).
Substituting $v^*_i(p_Q)$ in Eq.(\ref{maxM1}) we get
\beq
\z_Q(p_Q) = \z_W(p_Q-v^*_i(p_Q)) - \b \t^{-1} \log[\t v^*_i(p_Q)]
\sim - \frac{\b}{\t} \log[ \t (p_Q-1) ] + O( \t^{-1} )
\eeq
Note however that constant shifts of the order of $\log(\t)/\t$ are present
due to the factor $\t^2$ in front of Eq.(\ref{Pstart}) and from the Gaussian
fluctuations around the saddle-point. An empirical way to find the constant is
to impose the continuity of $\z_Q(p_Q)$, i.e. $\z_Q(p_+)=\z_W(p_+)$.
From this it follows
\beq
\z_Q(p_Q) = \z_W(p_Q-v^*_i(p_Q)) - \b \t^{-1} \log[2 v^*_i(p_Q)/\d]
- \z_W(1+\d/2) + \z_W(1+\d) \ .
\eeq
%An example is in Fig.~\ref{fig_2}.

%\begin{figure}
%\includegraphics[width=10cm]{zeta.eps}
%\caption{
%The function $\z(p)$ for $p>1$ in the power-law case. 
%Here $\z_0(p)=-(p-1)^2/4$ and $\b=6$.
%}
%\label{fig_2}
%\end{figure}

A similar calculation can be done for $p_Q<1$ and leads to
\beq
\z(p_Q) \sim - \frac{\b}{\t} \log[ \t |p_Q-1| ]
\eeq
for large $p_Q$.

\end{document}